\documentclass[namedreferences]{solarphysics}

\usepackage[hyperref,optionalrh]{spr-sola-addons} 
\usepackage{graphicx}        
\usepackage{amssymb}        
\usepackage{color}           
\usepackage{breakurl}        

\usepackage{xcolor}           

\renewcommand{\vec}[1]{{\mathbfit #1}}

\chardef\us=`\_

\begin{document}

\begin{article}
\begin{opening}

\title{Zebra stripes with high gyro-harmonic numbers}

\author[addressref={aff1},email={benacek@tu-berlin.de}]{Jan Ben\'a\v{c}ek \orcid{0000-0002-4319-8083}}
\author[addressref=aff2,email={karlicky@asu.cas.cz}]{Marian Karlick\'y \orcid{0000-0002-3963-8701}}

\address[id=aff1]{Center for Astronomy and Astrophysics, Technical University of Berlin, 10623 Berlin, Germany}
\address[id=aff2]{Astronomical Institute, Czech Academy of Sciences, 251 65 Ond\v{r}ejov,
    Czech Republic}

\runningauthor{Ben\'a\v{c}ek \& Karlick\'y}
\runningtitle{Zebra stripes with high gyro-harmonic numbers}

\begin{abstract}
Solar radio zebras are used in the determination of the plasma density and
magnetic field in solar flare plasmas. Analyzing observed zebra
stripes and assuming their generation by the double-plasma resonance (DPR)
instability, high values of the gyro-harmonic number are found. In some cases,
they exceed one hundred, in disagreement with the DPR growth rates computed up
to now, which decrease with increasing gyro-harmonic number. We address the
question of how the zebras with high values of the gyro-harmonic numbers $s$ are
generated. For this purpose, we compute growth rates of the DPR instability in
a very broad  range of $s$, considering a loss-cone $\kappa$-distribution of
superthermal electrons and varying the loss-cone angle, electron energies, and
background plasma temperature. We numerically calculated dispersion relations
and growth rates of the upper-hybrid waves and found that the growth rates
increase with increasing gyro-harmonic numbers if the loss-cone angles are
$\sim80^\circ$. The highest growth rates for these loss-cone angles are
obtained for the velocity $v_\kappa = 0.15\,c$. The growth rates as function
of the gyro-harmonic number still show well distinct peaks, which correspond to
zebra-stripe frequencies. The contrast of the growth rate peaks to
surrounding growth rate levels increases as the $\kappa$ index
increases and the background temperature decreases. Zebras with high values of
$s$ can be generated in regions where loss-cone distributions of
superthermal electrons with large loss-cone angles ($\sim80^\circ$) are
present. Furthermore, owing to the high values of $s$, the
magnetic field is relatively weak and has a small spatial gradient in such regions.
\end{abstract}
\keywords{Radio Bursts, Type IV; Radio Emission, Theory; Instabilities}
\end{opening}

\section{Introduction}
\label{sec:intro} 
\begin{figure}[!ht]
    \includegraphics[width=\textwidth]{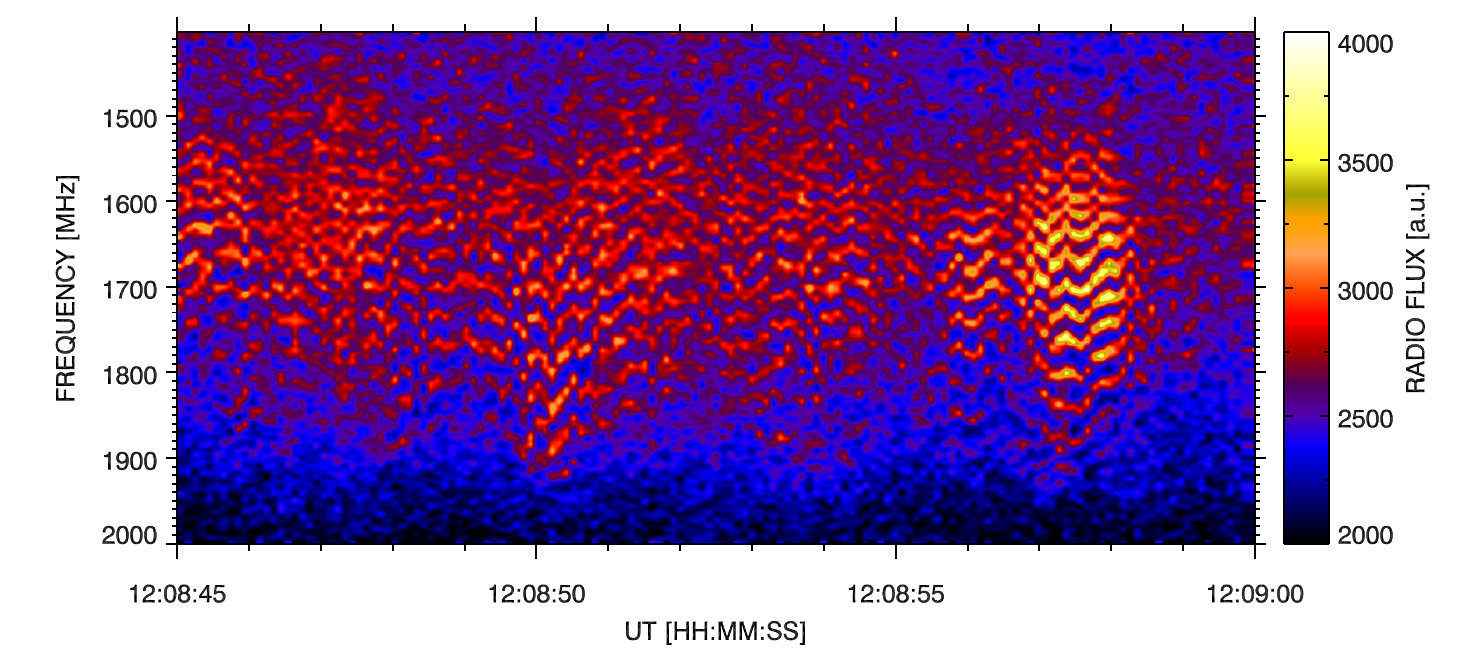}
    \caption{An example of the zebra observed at 14 February 1999 by the Ond\v{r}ejov
    radiospectrograph \citep{2008SoPh..253...95J}.}
    \label{fig0}
\end{figure}

Solar radio bursts and their fine structures are used to
diagnose solar flare plasmas. Among them, zebra patterns (zebras) belong to
the most frequently investigated bursts. An example of the zebra observed in 14
February 1999 is shown in Figure~\ref{fig0}. In the radio spectrum, they
appear in the form of regularly spaced emission stripes (zebra stripes). Many
papers and monographs were devoted to them, e.g.,
\citet{slo81,2006SSRv..127..195C,2011fssr.book.....C,2012ApJ...744..166T,2014ApJ...780..129T,Chernov2020}.

For the interpretation of the zebra stripes, several models have been proposed \citep{1972SoPh...25..188R,
    1975PhDT.........1K, 1975SoPh...44..461Z, 1976SvA....20..449C,
    1990SoPh..130...75C, 2003ApJ...593.1195L, 2007SoPh..241..127K,
    2006A&A...450..359B, 2006SoPh..233..129L, 2009PlPhR..35..160L,
    2010Ap&SS.325..251T, 2013A&A...552A..90K}.
In the published literature, the most often used model of zebras is the one based on the
double plasma resonance (DPR) condition
\citep{1975PhDT.........1K,1975SoPh...44..461Z,1980IAUS...86..341K,
    1983SoPh...83..305M,1988SoPh..116..323M,1986ApJ...307..808W,2013SoPh..284..579Z}
\begin{equation}
\label{eq1}
\omega - \frac{s \omega_\mathrm{ce}}{\gamma} - \frac{k_\parallel u_\parallel}{\gamma} = 0,
\end{equation}
where $\omega$ and $\vec{k} = (k_\parallel, k_\perp)$ are the frequency and
wave vector of the resonant wave, $\omega_\mathrm{ce}$ is the electron
cyclotron frequency, $s$ is the gyro-harmonic number, $\gamma = \sqrt{1 +
    (u_\parallel^2 + u_\perp^2)/c^2}$ is the relativistic Lorentz factor,
$u_\parallel = p_\parallel/m_\mathrm{e}$, $u_\perp = p_\perp/m_\mathrm{e}$, in the electron
momentum $\vec{p} = (p_\parallel,p_\perp)$, and $m_\mathrm{e}$ is the electron mass. This
condition may be simplified assuming that the generated waves are  upper-hybrid (UH) waves in a
non-relativistic plasma and the DPR resonance is at integer gyro-harmonics of
the electron-cyclotron frequency
\begin{equation}
f_{\rm UH} = (f_{\rm pe}^2 + f_{\rm ce}^2)^{1/2} \approx s f_{\rm ce},
\label{eq2}
\end{equation}
where $f_{\rm UH}$ and $f_{\rm pe}$ are the UH and electron-plasma frequencies,
$f_{\rm ce}$ is the electron-cyclotron frequency.
In the DPR model, superthermal electrons are considered as the driver of the zebra
emission. Those electrons are considered to be trapped in a flare loop
and have a loss-cone distribution. In the flare loop locations where such a distribution is present and
the DPR condition is fulfilled, the DPR instability generates the UH waves.
These UH waves are then converted to the observed electromagnetic waves by
merging with low-frequency waves or by scattering on plasma particles. This
model is supported by positional observations made by the Owens Valley Solar
Array \citep{2011ApJ...736...64C}.

When applying the DPR model to observed zebras and to estimate plasma
parameters at the zebra source, the most important is to determine the
gyro-harmonic numbers of zebra stripes $s$. From the zebra stripe frequencies
and their gyro-harmonic numbers, the electron cyclotron frequency and magnetic
fields are consequently estimated. Several methods for their determination were
developed, as shown, e.g., in
\cite{2015A&A...581A.115K,2020SoPh..295...96Y,2021SoPh..296..139Y}.

These methods are based on fitting a set of the observed
zebra-stripe frequencies $f(n)$ by the set of theoretically derived
zebra-stripe frequencies $f_\mathrm{th}(n)$ in the form \citep{2021SoPh..296..139Y}
\begin{equation}
f_\mathrm{th}(n)=\frac{(s_1-(n-1)) f(s_1)}{s_1} \left(\frac{(s_1-(n-1))^2-1}{s_1^2-1}\right)^{\frac{1}{R-2}},
\label{eq3}
\end{equation}
where $n$ is the ordinal number of the zebra stripes counted from the lowest to
higher frequencies, $f(s_1)$ is the frequency of the stripe with the lowest
frequency, $s_1$ is the corresponding gyro-harmonic number, $R=L_\mathrm{bh}$/$L_\mathrm{nh}$
is the ratio of spatial scales, which exponentialy depend on the magnetic
field and density, in the region where zebra-stripes are generated, and $R =
R_1 + R_2 (n - 1)$ is taken as a linear function of $n$, where $R_1$ and
$R_2$ are constants.

The fitting procedure is mathematically expressed as a search for the minimum value in the relation
\begin{eqnarray}
\frac{1}{N} \sum_{n=1}^{N} (f_\mathrm{th}(n) - f(n))^2,
\label{eq4}
\end{eqnarray}
where $N$ is the total number of the analyzed zebra stripes. The parameters $s_1,
R_1, R_2$, and $f(s_1)$ are the result of this fitting procedure. For more details
about this method, see \cite{2020SoPh..295...96Y} and \cite{2021SoPh..296..139Y}.

Using this method, seven detected zebras were analyzed at selected
times, and gyro-harmonic numbers $s_1$ were determined for the zebra-stripe
with the lowest analyzed frequency $f(s_1)$, see Table~\ref{tab1}. As can be
seen there, the gyro-harmonic number $s_1$ is greater than 50 in all these
zebras. It is surprising that the physical process that generates zebra stripes
operates on such high harmonics instead of the usually preferred low harmonics,
where the resonance frequency might be achieved easily. These gyro-harmonic
numbers are also significantly higher than those $s$ that have been considered
in previous theoretical studies of the DPR instability. Moreover, the
theoretical estimations made until now show that both growth rates and saturation
energies of the UH waves caused by the DPR instability decrease with increasing gyro-harmonic number \citep{2007SoPh..241..127K,2018A&A...611A..60B}. However,
the growth rates and saturation energies of the DPR instability have been so
far mainly computed for gyro-harmonic numbers $s \lesssim 20$
\citep{1986ApJ...307..808W,2017A&A...598A.106B,2018JGRA..123.7320L,Ni_2020,Li_Chen_Ni_Tan_Ning_Zhang_2021}.
 Only \citet{2007SoPh..241..127K}
calculated the growth rates for $s \lesssim 50$ for a power-law velocity
distribution with a loss-cone angle $30^\circ$, showing that the growth rate
maxima are well distinct for selected parameters.

In the present paper, we propose a solution to the problem of the
generation of the zebra stripes with high gyro-harmonic numbers
by considering a loss-cone distribution with large loss-cone angles.
We compute the growth rates of the DPR instability in a broad interval of UH to
cyclotron frequency ratios, corresponding to observed gyro-harmonic numbers up
to $s\sim$100. We solve this problem numerically, integrating the analytical
expressions for these growth rates. We consider the loss-cone
$\kappa$-distribution and vary the loss-cone angle, electron energies,
background plasma temperature, and the $\kappa$ parameter. Besides the
general results covering a broad range of gyro-harmonic numbers, we also present
some results for the zebra observed on the 14 February 1999.

The paper is organized as follows: In Section~2, we describe the method of the
growth rate calculation. Section~3 presents the results of these calculations
for a broad range of gyro-harmonic numbers. Finally, we summarize our results
and discuss how zebra stripes with high values of $s$ can be generated in Section~4.

\begin{table}
\centering
\begin{tabular}{r|c|c|c}
\hline\hline
Date & Time & f ($s_1$) & Gyro-harmonic \\
     & (UT) &  (MHz)    & number $s_1$ \\
\hline
25 Oct 1994 & 10:08:23.1 & 144.8 & 78 \\
17 Aug 1998 & 07:06:31.0 & 260.3 & 78 \\
14 Feb 1999 & 12:08:57.0 & 1526  & 64 \\
21 Apr 2002 & 01:45:49.5 & 2668  & 52 \\
1 Aug  2010 & 08:21:16.0 & 1120  & 114 \\
24 Feb 2011 & 07:41:08.8 & 2977  & 70 \\
21 Jun 2011 & 03:22:27.4 & 160.7 & 152 \\
\hline
\end{tabular}
\caption{Parameters of analyzed zebras according to
\cite{2021A&A...646A.179K}.} \label{tab1}
\end{table}

\section{Method for the Growth Rate Calculation}
We assume a collisionless plasma consisting of a cold background plasma and hot electron
 species that are uniformly distributed in space and embedded in a uniform magnetic field.
The background plasma consists of cold electrons with number density $n_\mathrm{b}$
 and cold protons with density $n_\mathrm{i}$.
Hot electrons of density $n_\mathrm{h}$ are also considered.
The plasma satisfies the charge neutrality condition, $n_\mathrm{i} = n_\mathrm{b} + n_\mathrm{h}$.
Both cold particle species have Maxwell velocity distributions with thermal velocities
 $v_\mathrm{b,i} = \sqrt{k_\mathrm{B} T_\mathrm{b,i} / m_\mathrm{b,i}}$,
  where $k_\mathrm{B}$ is the Boltzmann constant, $T_\mathrm{b,i}$ are the thermodynamic
   temperatures, and $m_\mathrm{b,i}$ are the masses of cold electrons and protons, respectively.

The hot electrons have a loss-cone $\kappa$-distribution that can be expressed in
cylindrical coordinates as
\begin{equation}
f(\vec{u}, \theta_\mathrm{c}, \Delta \theta_\mathrm{c}, \kappa) = f_\kappa(\vec{u}, \kappa) \, \varphi(\theta_\mathrm{c}, \Delta \theta_\mathrm{c}),
\end{equation}
where $\varphi$ characterizes the linear transition at loss cone angles
\begin{equation}
\varphi(\theta_\mathrm{c}, \Delta \theta_\mathrm{c}) =
\left\{
\begin{array}{ll}
0,  & ~~~~~~ \theta_\mathrm{c} - \Delta \theta_\mathrm{c} \geq \theta,  \\
\frac{\theta - (\theta_\mathrm{c} - \Delta \theta_\mathrm{c}) }{2 \Delta \theta_\mathrm{c}}, & ~~~~~~ \theta_\mathrm{c} + \Delta \theta_\mathrm{c} > \theta > \theta_\mathrm{c} - \Delta \theta_\mathrm{c}, \\
1,  & \pi - \theta_\mathrm{c} - \Delta \theta_\mathrm{c}  \geq \theta \geq \theta_\mathrm{c} + \Delta \theta_\mathrm{c}, \\
\frac{(\pi - \theta_\mathrm{c} + \Delta \theta_\mathrm{c}) - \theta }{2 \Delta \theta_\mathrm{c}}, & \pi - \theta_\mathrm{c} + \Delta \theta_\mathrm{c} > \theta > \pi - \theta_\mathrm{c} - \Delta \theta_\mathrm{c}, \\
0,  & ~~~~~~~~~~~~~~~~~~~~~~~~~\theta \geq \pi - \theta_\mathrm{c} + \Delta \theta_\mathrm{c}, \\
\end{array}
\right.
\end{equation}
where $\theta_\mathrm{c}$ is the loss-cone angle,
$\Delta \theta_\mathrm{c}$ is the angular half-width of the linear loss-cone transition,
$\theta = \arctan(u_\perp/u_\parallel)$ is the particle pitch angle,
$f_\kappa(\vec{u}, \kappa)$ is the $\kappa$-distribution \citep{2013SSRv..175..183L}
\begin{equation}
f_\kappa(\vec{u}, \kappa) = \frac{n_\mathrm{h}}{n_\mathrm{e}}\frac{1}{(\pi \kappa v_\mathrm{\kappa}^2 )^{\frac{3}{2}} \cos(\theta_\mathrm{c})} \frac{\Gamma (\kappa + 1)}{\Gamma( \kappa - \frac{1}{2})} \left( 1 + \frac{u_\perp^2 + u_\parallel^2}{\kappa v_\mathrm{\kappa}^2} \right)^{- \kappa -1},
\end{equation}
characterized by the index $\kappa = \left(\frac{3}{2}, \infty\right)$,
$v_\mathrm{\kappa}$ is the $\kappa$-distribution velocity,
and $\Gamma$ is the Gamma function. The
distribution function is normalized to the loss-cone angle as 
$1/\cos(\theta_\mathrm{c})$, assuming that $\Delta \theta_\mathrm{c} \ll 1$.
Unless otherwise mentioned, we use $\kappa = 2$ and $n_\mathrm{e} /
n_\mathrm{h} = 32$. The $\kappa$ index is in the interval of the observed
values during solar flares, see, e.g., \citet{2020ApJ...893...34L}. The loss-cone
angular half-width is taken as $\Delta \theta_\mathrm{c} = 2.5^\circ$.

The growth rate of the DPR instability is computed as
\citep{Melrose1986}
\begin{equation} \label{eq:growth-rate}
\gamma(\omega, k_\perp) = - \frac{ \mathrm{Im} \, \epsilon_\parallel^{(1)}}
{\left[\frac{\partial \mathrm{Re} \, \epsilon_\parallel^{(0)}}
    {\partial \omega}\right]_{\epsilon_\parallel^{(0)} = 0}},
\end{equation}
where $\mathrm{Re} \, \epsilon_\parallel^{(0)}$ and $\mathrm{Im} \, \epsilon_\parallel^{(1)}$
 are the real and imaginary parts of the plasma dispersion tensor component along the magnetic
  field, that are obtained by a perturbation approach from the total dispersion tensor component
\begin{equation}
\epsilon_\parallel = \epsilon_\parallel^{(0)} + \epsilon_\parallel^{(1)} + \mathcal{O}(\epsilon_\parallel^{(2)}).
\end{equation}
The imaginary part of the tensor component in
Equation~\ref{eq:growth-rate} for the growth rate was derived by
\citet{2005A&A...438..341K}
\begin{eqnarray}
\mathrm{Im}(\epsilon_\parallel^{(1)}) &=& - 2\pi^2 m_\mathrm{e}^3  \frac{\omega_\mathrm{pe}^2}{k^2}
\sum_{l=s+1}^{\infty} a^3 \times \nonumber \\
&& \times \int_0^{\pi}  J_l \left( \frac{\gamma_\mathrm{rel} k_\perp v_\perp}{\omega_\mathrm{ce}} \right) \frac{\gamma_\mathrm{rel}^2 \sin \phi}{\frac{\partial \psi}{\partial\rho}}  \frac{1}{ v_\perp}\frac{\partial f}{\partial u_\perp} \, \mathrm{d}\phi,
\label{eq-growth}
\end{eqnarray}

\begin{equation}
\frac{\partial \psi}{\partial \rho} = \frac{1}{c^2} \left( v_\parallel^2 + v_\perp^2 \right).
\end{equation}

\begin{equation}
v_\parallel = - a \cos(\phi), \quad v_\perp = a \sin(\phi), \quad
a^2 = \frac{ c^2 (l^2 \omega_\mathrm{ce}^2 - \omega^2)}
{l^2 \omega_\mathrm{ce}^2}
\end{equation}
\begin{equation} \label{eq:lambda}
\omega_\mathrm{pe}^2 = \omega_\mathrm{pb}^2 + \omega_\mathrm{ph}^2 = \frac{ (n_\mathrm{b} + n_\mathrm{h}) e^2}{m_\mathrm{e}
    \epsilon_\mathrm{0}}, \qquad \lambda = \frac{k_\perp^2
    v_\mathrm{tb}^2}{\omega_\mathrm{ce}^2},
\end{equation}
where $\omega_\mathrm{pe}$ is the plasma frequency of both electron species,
$\omega_\mathrm{ce} = e B / m_\mathrm{e}$ is the electron cyclotron frequency,
$\mathbf{k} = (k_\perp, k_\parallel)$ is the wave vector of the electrostatic waves and
we assume $k_\parallel = 0$, $\omega$ is the wave frequency,
$I_l(\lambda)$ is the modified Bessel function of the first kind of the $l$-th order,
$\lambda$ is the dimensionless parameter of the Bessel function,
$e$ is the electron charge,
$\epsilon_0$ is the permittivity of vacuum,
and $s$ is the gyro-harmonic number.
The integration is over the resonance ellipse of Equation~\ref{eq1}.

The denominator in Equation~\ref{eq:growth-rate} can be expressed as
\begin{equation} \label{eq:partial_epsilon_omega}
\frac{\partial \epsilon_\parallel^{(0)}}{\partial \omega} = 4 \omega \omega_\mathrm{pb}^2 \frac{e^{-\lambda}}{\lambda} \sum_{l=1}^\infty \frac{l^2 I_l(\lambda)}{(\omega^2 - l^2 \omega_\mathrm{ce}^2)^2}.
\end{equation}
Equation~\ref{eq:growth-rate} can be mathematically evaluated in the
$\omega-k_\perp$ domain; however, only the waves that are solutions of the
plasma dispersion relation are actually present in the plasma and can grow. To
simplify the calculation of growth rates, we choose the dispersion
relation of the UH waves as
\begin{equation} \label{eq:UH-waves}
\omega^2 = \omega_\mathrm{pb}^2 + \omega_\mathrm{ce}^2 + 3 v_\mathrm{tb}^2 k_\perp^2,
\end{equation}
where we assume that the dispersion relation is given only by the background plasma component,
 $\epsilon_\parallel^{(0)}$.

The numerical procedure for the computation of a growth rate value for a given set
of plasma parameters ($\omega_\mathrm{pe}/\omega_\mathrm{ce}$, $v_\mathrm{tb}$,
$v_\mathrm{\kappa}$, $\kappa$, $\Delta \theta_\mathrm{c}$, and
$\theta_\mathrm{c}$) is the following. We equidistantly divide the wavenumber
space of the UH branch into $N=1500$ grid points in
the $k_\perp c /\omega_\mathrm{pe} \in [1,40]$ interval.
According to our tests, this interval is wide enough for the whole range of
studied background temperatures.
Approximately, the typical wavenumber $k_\perp$ increases with decreasing background thermal
velocity $[v_\mathrm{tb}]$ as follows from Equation~\ref{eq:lambda} when the parameter
 $\lambda$ is constant. The unstable waves are typically located in an
interval $k_\perp c / \omega_\mathrm{pe} \approx [4,15]$ for a temperature of 3\,MK
and in an interval $k_\perp c / \omega_\mathrm{pe} \approx [10,37]$ for a
temperature of 0.5\,MK. Moreover, these intervals slightly vary with other parameters.
The grid cell size is $\Delta k_\perp c /\omega_\mathrm{pe} = 0.026$. For each
grid cell $k_{\perp i}$, where $i=1, \ldots, 1500$ is the grid cell index, we compute the
frequency of the UH wave $\omega_i$ from Equation~\ref{eq:UH-waves}. For the
frequency [$\omega_i$], we
evaluate the growth rate $\gamma_i$ in each grid cell [$k_{\perp i}$]. Then, we compute
the mean value of the growth rate over all
positive $\gamma_i$ as
\begin{equation} \label{eq:integrated}
\Gamma = \frac{1}{\Gamma_0} \sum_i \gamma_i(\omega_i, k_{\perp i}) \sigma_i(\omega_i, k_{\perp i}) \Delta k_{\perp},
\end{equation}
where $\sigma_i$ is the characteristic width of the dispersion branch for each $i$-th element
\citep{2019ApJ...881...21B,2021A&A...649A.145M}.
The characteristic width $\sigma_i$ is estimated as
\begin{equation}
\sigma_i = \left. \left(\frac{\partial \epsilon_\parallel^{(0)}}{\partial \omega} \right)^{-1} \right|_{\omega_i,k_{\perp i}},
\end{equation}
$\Gamma_0$ is the normalization parameter
\begin{equation}
\Gamma_0 = \sum_i \sigma_i(\omega, k_\perp) \Delta k_{\perp},
\end{equation}
We denote the mean growth rate value $\Gamma$ as \textit{integrated growth rate}.

\begin{figure*}[!ht]
    \includegraphics[width=\textwidth]{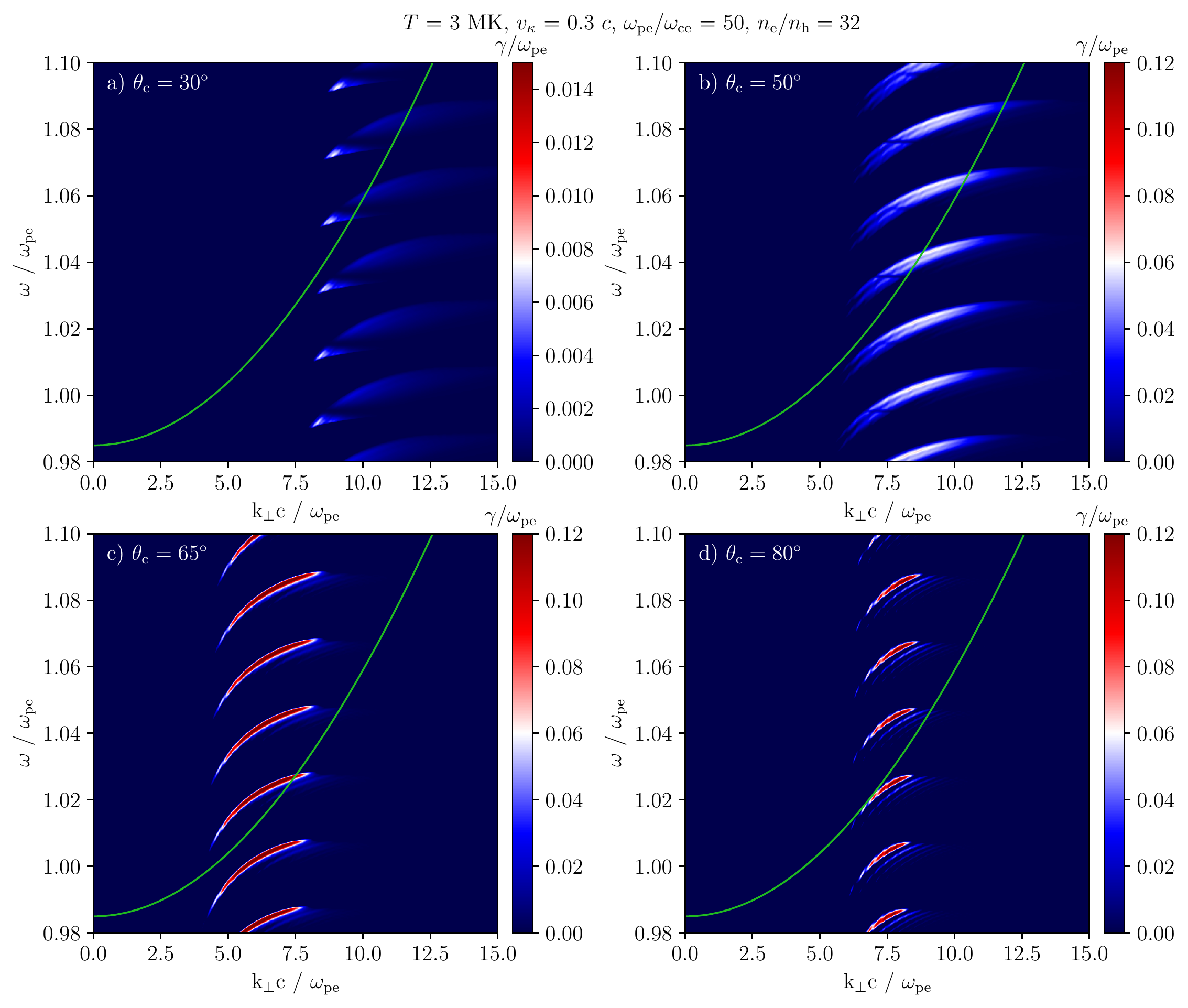}
    \caption{Growth rates as a function of the frequency and wavenumber for four loss-cone
        angles $\theta_\mathrm{c}$, the gyro-harmonic number is $s = 50$, the temperature
        of the background plasma 3\,MK, and the velocity of $\kappa$-distribution
        $v_\mathrm{\kappa} = 0.3\,c$.
        Note that the color scale in (a) differs from those in (b-d).
        \textit{Green curves:} the dispersion branch of UH waves. All these UH branches are the same
        in all panels because they are independent of the loss-cone distribution.}
    \label{fig1}
\end{figure*}

\section{Results} 
\label{sec:results}
In the following, our results on the growth rate
of the electrostatic waves are shown as a function of the gyro-harmonic number $s =
\sqrt{\omega_\mathrm{pe}^2 / \omega_\mathrm{ce}^2 + 1}$, as follows from the
radio emission process. Because the frequency ratio changes with the height in the
corona, zebra stripes are produced only when the profile
of integrated growth rate forms peaks at specific frequency ratios. The peaks
are not always located at integers of $s$ (harmonics of the
cyclotron frequency) because there are other effects that can shift them,
such as the effect due to the Lorentz factor in Equation~\ref{eq1}, as was
shown by \citet{2017A&A...598A.106B}.

Figure~\ref{fig1} shows growth rates in the $\omega-k_\perp$ space for four
selected loss-cone angles $\theta_\mathrm{c}$ = 30$^{\circ}$, 50$^{\circ}$,
65$^{\circ}$, and 80$^{\circ}$, the background temperature 3\,MK
($\approx$130\,eV), and the velocity of $\kappa$-distribution $v_{\kappa} = 0.3$\,c
($\approx$23\,keV). The gyro-harmonic number is fixed to $s = 50$. The growth
rates are smallest for the small loss-cone angle $30^\circ$.
By increasing the loss-cone angles to $50^\circ$ and $65^\circ$, the growth rate values
increase and the positive growth rate regions shift to smaller wavenumbers.
For the loss-cone angle $80^\circ$, the growth rates values and areas
of positive growth rate regions decrease.

The growth rates are overlaid by the dispersion branch of the UH waves
(Equation~\ref{eq:UH-waves}). If the branch crosses a positive growth rate
region, the waves located in the intersection can grow.
During the growth phase, the kinetic energy of resonant
hot electrons, whose velocities fulfill Equation~\ref{eq1}, is converted into
electrostatic wave energy and the particle distribution changes. Finally, the
growing waves saturate when the electron velocity distribution becomes stable.

Each region in the $\omega-k_\perp$ space  with a positive growth
rate is associated with a specific gyro-harmonic number $s$.
These regions have a distance in frequency dimension approximately equal
to the electron cyclotron frequency.
When the frequency ratio varies, the positive regions shift in frequency so that
the crossings of positive growth rate regions with the UH branch change. This way we expect
the formation of growth rate peaks (which may be responsible for the formation
of zebra stripes). However, the peaks are not so
significant if the UH branch crosses several positive regions simultaneously
(e.g., Figure~\ref{fig1}(b)).

\begin{figure*}
    \includegraphics[width=\textwidth]{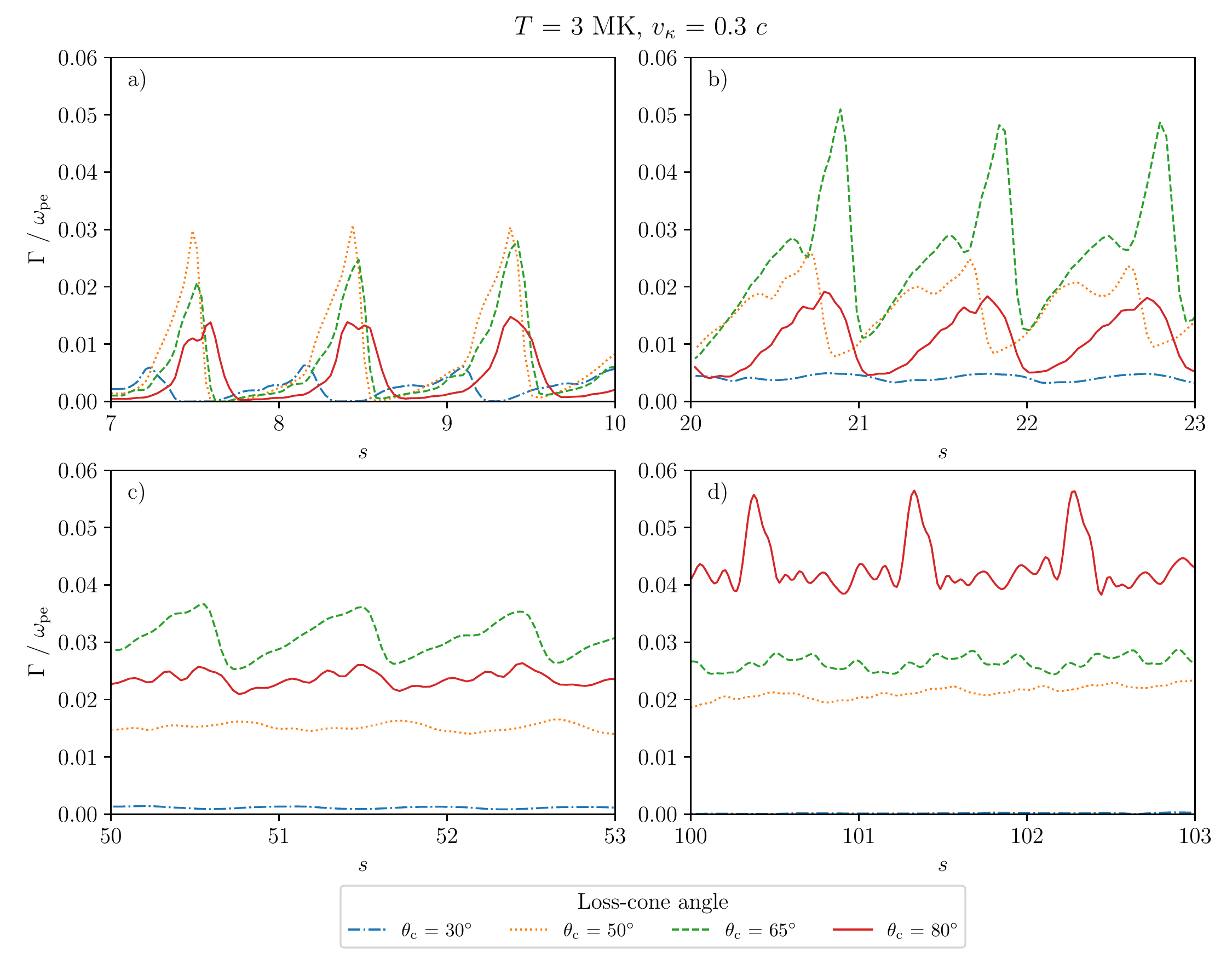}
    \caption{Integrated growth rates in four selected intervals of
        the gyro-harmonic numbers and for four loss-cone angles. The temperature of the
        background plasma is 3\,MK,
        and the velocity of the $\kappa$-distribution $v_\mathrm{\kappa} = 0.3\,c$.}
    \label{fig2}
\end{figure*}

In Figure~\ref{fig2}, the integrated growth rates are presented
as a function of the gyro-harmonic number for four loss-cone angles, background
temperature 3\,MK, and the velocity of the $\kappa$-distribution $v_\kappa = 0.3\,c$. The
integrated growth rates are computed following
Equation~\ref{eq:integrated}. We selected four intervals of gyro-harmonic numbers
from small values $s = 7$, for which the radio zebras are not observed, up to
the most extreme case of the gyro-harmonic number exceeding 100. In all the
cases the growth rates are positive.
For low gyro-harmonic numbers $s = 7-10$, the growth rates form
distinguishable peaks for all loss-cone angles.  By increasing $s$
the growth rate profiles smooth out; however, the smoothing varies
between different loss-cone angles. For the angle 30$^\circ$, the growth rate
profile smooths out already for $s \gtrsim 20$, and their values decrease for
higher gyro-harmonic numbers. For the angles 50$^\circ$ and 65$^\circ$, the
growth rate profiles smooth out in the gyro-harmonic number interval $s =
50-53$. Their average values slightly increase and decrease for
$\theta_\mathrm{c} = 50^\circ$ and for $\theta_\mathrm{c} = 65^\circ$,
respectively. For the loss-cone angle 80$^\circ$, the growth rates monotonically increase
within the whole studied interval of gyro-harmonic numbers. At large
gyro-harmonic numbers $s = 100-103$, the growth rate peaks are still formed.

\begin{figure*}
    \includegraphics[width=\textwidth]{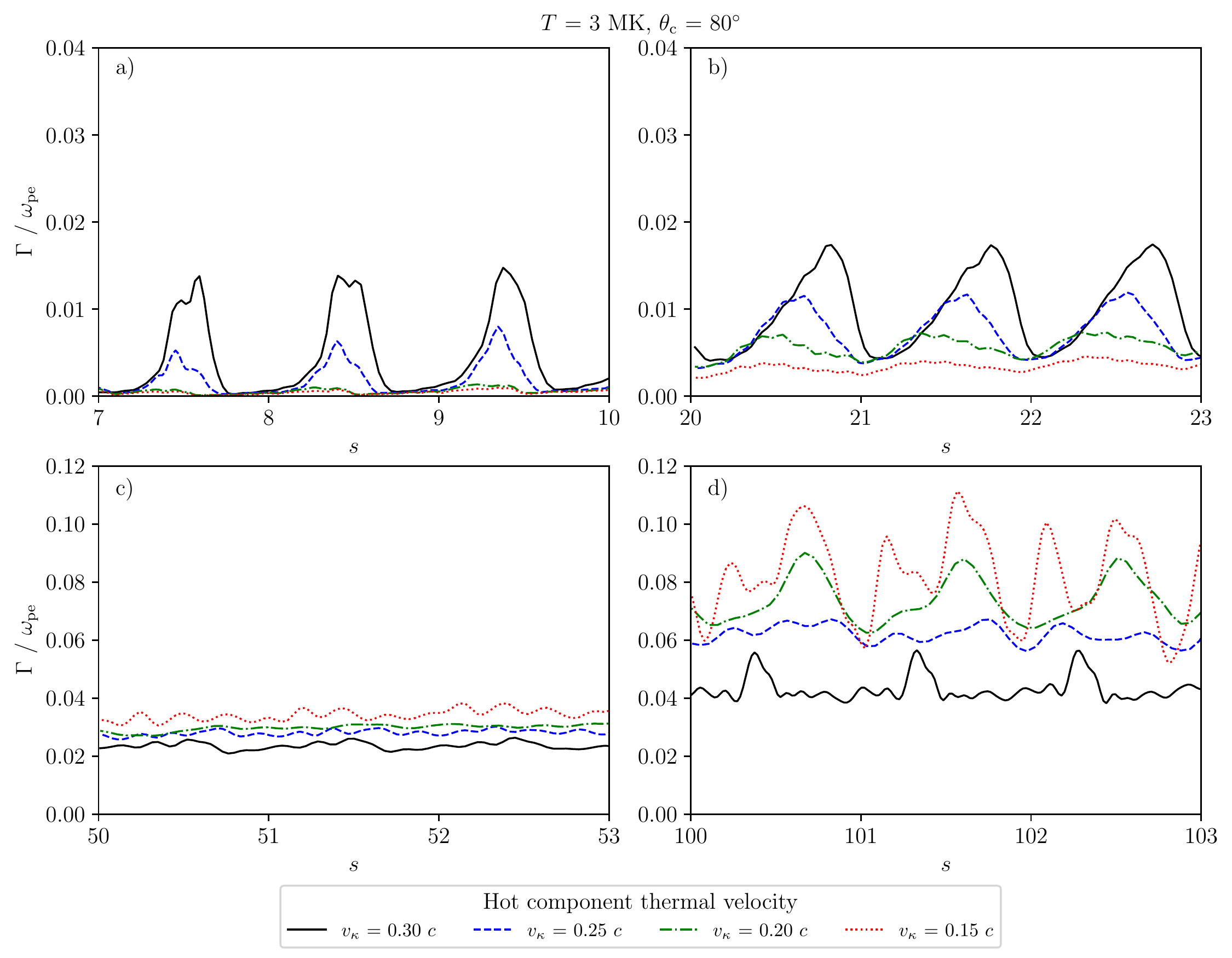}
    \caption{Integrated growth rates as a functions of the $\kappa$-distribution velocity,
        for four selected intervals of the gyro-harmonic number, the loss-cone angle 80\,$^{\circ}$
        and the temperature of the background plasma 3\,MK. Note that the scaling of the
        growth rates varies between panels.}
    \label{fig3}
\end{figure*}

\begin{figure*}
    \includegraphics[width=\textwidth]{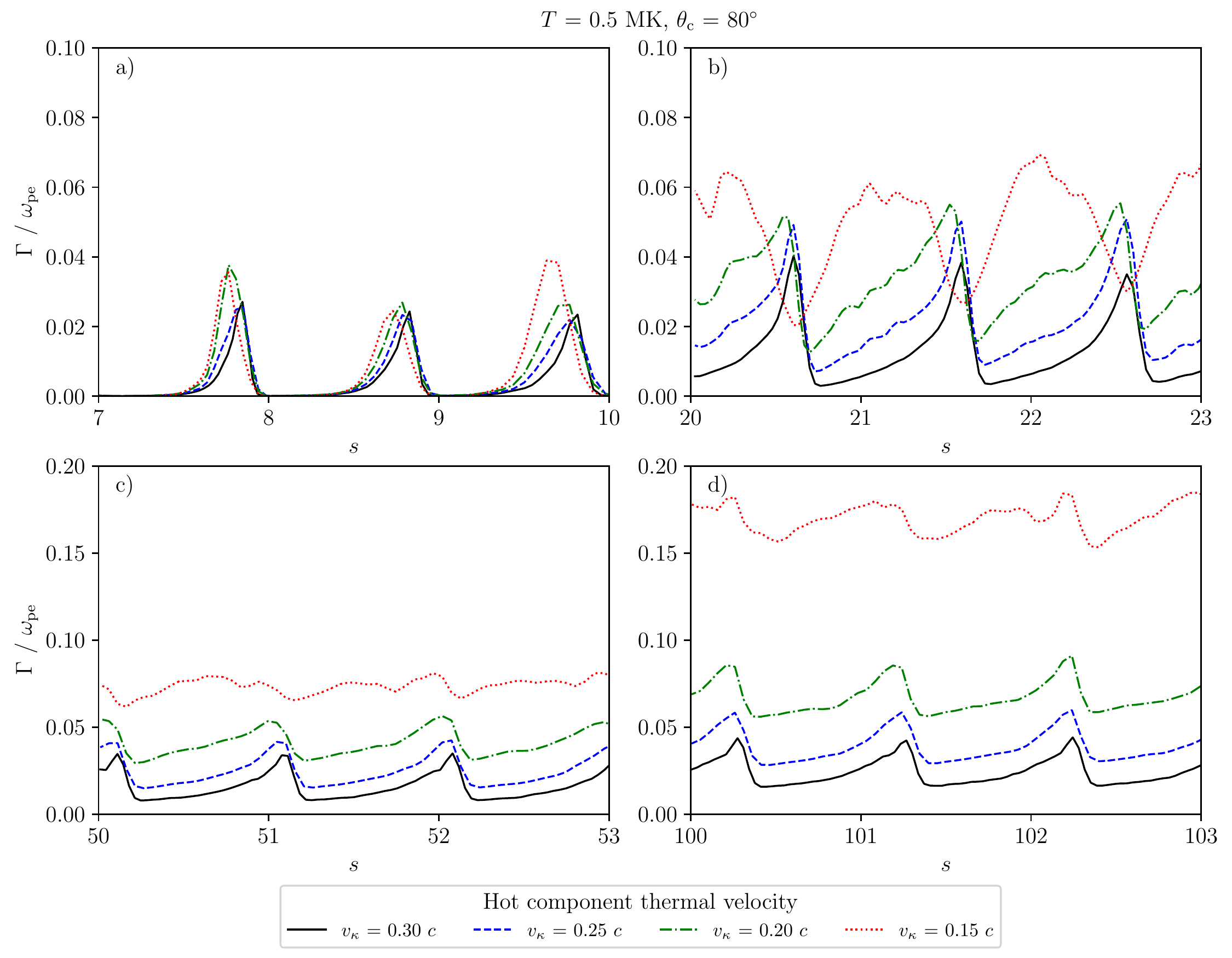}
    \caption{Same as Figure~\ref{fig3}, but for the temperature of the background plasma 0.5\,MK.}
    \label{fig4}
\end{figure*}

Because the highest growth rates for the large gyro-harmonic numbers
(derived from zebra observations) occur for the angle $80^\circ$, we
studied the dependence of integrated growth rate profiles on two plasma
temperatures (0.5 and 3~MK) and selected values for the velocities of the $\kappa$-distribution in the interval $v_\kappa \in [0.15,0.3]c$,
see Figures~\ref{fig3} and \ref{fig4}. Though the temperature 0.5~MK
may be smaller than those estimated during flares, the accelerated electrons may
reach and emit in the colder part of the corona. Moreover, we selected this background
temperature range to clearly identify the temperature impact as the
background temperature is known to have a small impact on the instability
\citep{2017A&A...598A.106B}. The integrated growth rates are computed in the
same gyro-harmonic number intervals as in Figure~\ref{fig2}.

For the plasma temperature 3~MK (Figures~\ref{fig3}), the average values (over few $s$) of
the integrated growth rates increase with the gyro-harmonic number independently on
the velocity of the $\kappa$-distribution. All growth rate profiles smooth out for
$s\sim50$. The situation is reversed for large gyro-harmonic numbers where
the most distinct peaks are formed for the smallest velocity of the $\kappa$-distribution
$v_\mathrm{\kappa} = 0.15$. In addition, the integrated growth rates
for this background velocity and gyro-harmonic numbers form a double peak
profile.

The behavior of the integrated growth rates as a function of the gyro-harmonic
number is different for the background temperature
$0.5$~MK in a few aspects, see Figure~\ref{fig4}. For velocities of the $\kappa$-distribution
$v_\mathrm{\kappa} \geq 0.2\,c$, the growth rate peaks are well distinct for
the whole range of the gyro-harmonic numbers. The distinction occurs
because the distribution
functions of the hot and cold components are more separated in velocity space
than for the temperature 3\,MK. Because of this separation, the velocity
gradient of the loss-cone distribution increases, and the growth rate
reaches higher values. For all velocities of the $\kappa$-distribution, the
growth rate values increase from $s = 7-10$ to $s = 20-23$. By further
increasing the gyro-harmonic numbers, the growth rate increases the most for the
smallest velocities of the $\kappa$-distribution. For the case of $v_\mathrm{\kappa} =
0.3\,c$, the growth rates remain almost the same as for $s
\gtrsim 20$.

\begin{figure}
    \centering
    \includegraphics[width=0.6\textwidth]{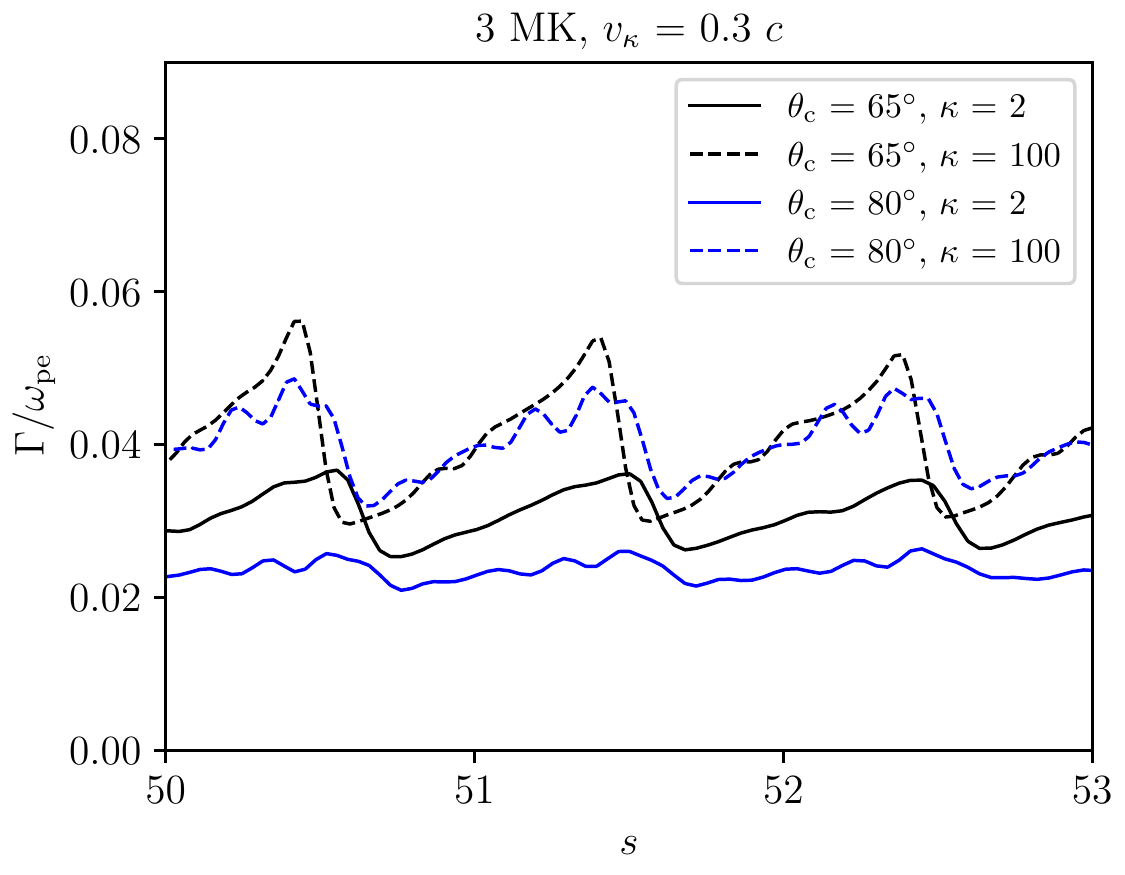}
    \caption{Comparison of the integrated growth rates for $\kappa$-indices 2 and 100,
        temperature 3~MK, and the velocity of the $\kappa$-distribution $0.3\,c$.}
    \label{fig5}
\end{figure}

Figure~\ref{fig5} shows a change in the integrated growth rates with the
$\kappa$ index. We compare integrated growth rates for $\kappa = 2$ and $\kappa
= 100$ (closer to the Maxwellian velocity distribution).
The integrated growth rate profiles for $\kappa = 100$ are
higher and with more distinct peaks than those for $\kappa=2$.

\subsection{The 14 February 1999 Zebra Case}

Besides the above presented general results covering a broad range of the
gy\-ro-har\-mo\-nic number, let us summarize the results for the
specific zebra observed on the 14th February 1999 and shown in Figure~\ref{fig0}, as
an example. In this case the gyro-harmonic number was estimated as $s_1$ = 64
(Table~\ref{tab1}). For this zebra, we calculated the integrated growth rates
in the range $s = 64-67$, the loss-cone angles $55-80\,^\circ$, the
temperatures 0.5\,MK and 3\,MK, and $v_\kappa \in [0.1,0.5]c$. From the analysis
follows that the emission is most likely produced by a loss-cone angle
$\sim (70\pm5)^{\circ}$. The growth rate peaks are formed for a wider
range of plasma temperatures and velocities of the $\kappa$-distribution.
For the plasma temperature of 3\,MK, the velocity of the $\kappa$-distribution must
be $\geq$ 0.15\,$c$. For the plasma temperature of 0.5\,MK, the velocity of the
$\kappa$-distribution can be in the whole analyzed range of $v_\kappa \in [0.1,0.5]c$. Our
conclusion for this zebra is that though the plasma temperature and the
velocity of the $\kappa$-distribution can be in a broad range of values to produce
the emission, the loss-cone angle should be restricted into a narrow range of
about $\sim(70\pm5)^{\circ}$.

\section{Discussion and Conclusions}
\label{sec:discussion} The solar radio zebras are used for diagnostics of
plasma parameters during solar flares. One of the crucial parameters is the
gyro-harmonic number in the emission region, which allows to estimate the upper-hybrid
to cyclotron frequency ratio. Though the gyro-harmonic numbers
determined from zebra observations are greater than about 50 and may exceed
one hundred, most of the studies have so far focused on significantly smaller
gyro-harmonic numbers, rarely higher than 20. We addressed the question of
how the solar radio zebras can be produced with such high gyro-harmonic
numbers by analyzing the DPR instability in a broad range of gyro-harmonic
numbers $s = 7-103$.

We calculated linear growth rates and studied the effects of the loss-cone angle,
velocity of the $\kappa$-distribution, the background plasma temperature, and the $\kappa$
index on the instability.
We found that the zebras with high gyro-harmonic numbers are produced by loss-cone
distributions with large loss-cone angles. Moreover, the higher the gyro-harmonic
number is, the larger the loss-cone angle needs to be to produce the instability.
For the gyro-harmonic number $\sim 100$, a loss-cone angle $\sim80^{\circ}$ is sufficient.
However, for even larger gyro-harmonic numbers, we expect that the necessary high angles
of the loss-cone will be hard to produce and therefore unlikely to be observed.

\subsection{Growth Rate Properties}

Our calculations show that for increasing gyro-harmonic number the
integrated growth rates of the DPR instability increase for large
values of the loss-cone angle ($\sim80^\circ$). Moreover, the growth rate
peaks are distinct for gyro-harmonic numbers $s\sim100$ and smooth for
smaller loss-cone angles $<80^\circ$. The integrated growth rates are large
enough for all loss-cone angles $\gtrsim50^\circ$ to produce zebra type of
radio emission.

In some cases, the growth rates show only small variations depending
on $s$. However, we note that these variations may produce a
significant difference in the saturation energy levels, as was shown using
particle-in-cell simulations by \citet{2018A&A...611A..60B}.

The growth rate average values (over a broader range of gyro-harmonic numbers)
monotonously decrease only for a loss-cone angle 30$^\circ$. In contrast,
for the loss-cone angle 80$^\circ$, the average values
mo\-no\-to\-nou\-sly increase.

Moreover, how distinct the integrated growth rate maxima are for high
gyro-harmonic numbers $s \sim 50-100$ and for large loss-cone angles depends on the
temperature of the background plasma and the velocity of the $\kappa$-distribution.
Assuming that the instability occurs in a heated plasma during a flare with
temperature 3~MK, the growth rate profiles are smoothed out around $s \sim 50$ but
form emission peaks for $s \sim 100$. Moreover, for the gyro-harmonic number
$s \sim 100$, they produce double-peaked profiles for small velocities of
the $\kappa$-distribution $v_\kappa = 0.15c$. If the
emission is formed in a region with a colder plasma of temperature
0.5~MK, e.g., further from the flare region, the growth rates are distinct for
the whole range of gyro-harmonic numbers.
Nonetheless, for both temperatures and large gyro-harmonic numbers, higher growth rates are favored for smaller $\kappa$-distribution velocities in the range $v_\kappa \in [0.15,0.20]c$ than for larger values in the range $v_\kappa \in [0.25,0.3]c$.

\subsection{Proposed Zebra Emission Regions}

The fact that high gyro-harmonic numbers
are detected in zebras means that the zebra-stripe sources are in regions where
the ratio of $\omega_\mathrm{pe}/ \omega_\mathrm{ce}$ is high.
In combination with the result that higher growth rates are favored for large
loss-cone angles, two possible explanations can be proposed:

\begin{enumerate}
    \item We may assume that the loop cross-section is only slightly wider at their
    midpoints than at their footpoints, in agreement with \cite{2000SoPh..193...53K}.
    In such a loop, the change of the magnetic field along the loop is
    small, and only superthermal electrons with high pitch angles can be
    trapped \citep{2020ApJ...894..158K}. This way, a loss-cone
    distribution with a high loss-cone angle is formed. Thus, the instability can
    generate the UH waves and the corresponding zebra stripes with high
    gyro-harmonic numbers. Moreover, for the zebra stripes with high $s$,
    the magnetic field in this loop needs to be relatively small
    ($\omega_\mathrm{pe}/ \omega_\mathrm{ce}\gg$1). We also note that the
    small change of the magnetic field strength along the loop agrees with the small
    difference in magnetic field strength derived from the neighboring
    zebra-stripe frequencies for high values of the gyro-harmonic numbers.
    Typical intensity of the magnetic field in these regions can be derived as $B$ = $f_s$/(2.8$\cdot s$)), where $B$ is the magnetic
    field strength in a unit of gauss, and f$_s$ is the zebra-stripe frequency in a unit of MHz
    with the gyro-harmonic number $s$ \citep{2018ApJ...867...28K}.
    For low values of $s$, these differences are much higher.
    Therefore, in a loop with a small magnetic field gradient,
    spatial distances of zebra-stripe sources with low values
    of $s$ will be much longer than those for high values of $s$.

    \item Considering the standard flare model \citep{2004psci.book.....A},
    a region with high values of $\omega_\mathrm{pe}/ \omega_\mathrm{ce}$ can
    be found below the X-point of the flare. The
    plasma density here is relatively high (comparable to the current sheet
    density), and the magnetic field in the reconnection outflow is
    relatively low. The electrons from reconnection move downward and
    penetrate into newly formed magnetic loops where they may be trapped
    \citep{1987SoPh..108..237S}. If the electrons initially had
    perpendicular velocities to the magnetic field larger than their
    parallel ones, they can form loss-cone distributions with large
    loss-cone angles. Electrons are accelerated not only at the X-point of
    magnetic reconnection but also during the shrinking of magnetic loops
    in the reconnection outflow \cite{2004A&A...419.1159K}.

\end{enumerate}

\subsection{Double-Peaked Pattern of Growth Rates}
We searched for the double-peaked pattern from Figure~\ref{fig3}(d), whether it
also occurs for smaller gyro-harmonic numbers, various temperatures, and
$\kappa$-dis\-tri\-bu\-tion velocities. But it did not appear anywhere in our
calculations. Also, we performed tests to discard the numerical uncertainty
at the origin of the double-peaked profiles.
We varied the ranges of calculations and the
precision of the numerical integrals and tested individual parts of
calculations for these parameters, but we found no discrepancies or errors.
Nonetheless, due to the complexity of the calculations, we could not say
whether there are hidden numerical artifacts produced by an approximation made in the
analytical growth rate derivation.

The double-peaked growth rate profile is generally formed when the growth
rate region in the $\omega-k_\perp$ space is not as the one shown in Figure~\ref{fig1},
i.e., when it is not forming only one tadpole region for each gyro-harmonic number,
but it forms two (or more) separated regions. Therefore, the UH branch
could cross two regions and create two growth rate maxima. For this happen,
it would however require a specific kind of velocity distribution function,
probably with two distinct regions in the $\omega-k$ space of positive
velocity gradients.

A question arises about how we could observationally recognize that the adjacent
zebra stripes are formed by double-peaked growth rates. Assuming that there are
two distinct regions in the $\omega-k$ space, they probably do not have the
same shape. Therefore, both generated zebra stripes could be created at
different emission intensities, with different frequency ratios between odd and
even stripes, and odd and even stripes might have different emission profiles
(e.g., odd stripes more narrow and even stripes broader). Nonetheless, if the
regions are similar and regularly distributed in the $\omega-k$ space,
also the emission profiles can be
similar. Moreover, we may speculate that a change or evolution of the
velocity distribution during the emission might allow transition from
a double-peaked to a single peaked profile, or vice versa, as in the
observed zebras with merging zebra stripes \citep{2005A&A...437.1047C}.

\subsection{Comparison with other Procedures of the Growth Rates Calculation}

\begin{figure*}[!h]
    \includegraphics[width=\textwidth]{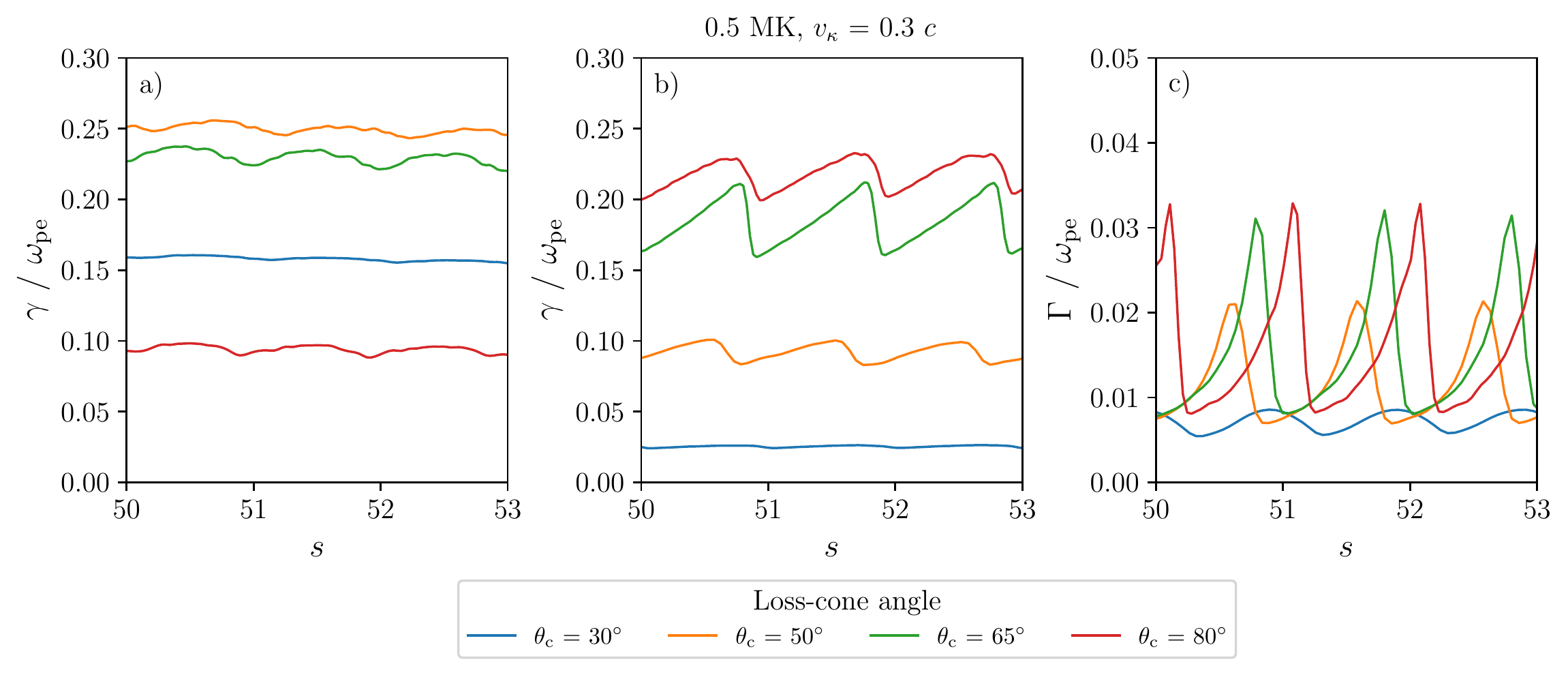}
    \caption{Comparison of the growth rate calculations: (a) using the maximal
        method of \citet{2007SoPh..241..127K} with (b) the maximal and (c)
        averaged method used in this paper and based on an integration over the
        whole resonance ellipse \citep{2005A&A...438..341K}.
        The background temperature is 0.5~MK, and the $\kappa$-distribution
        velocity is $v_\mathrm{\kappa} = 0.3\,c$.
        Compare the growth rates in panel (c) with the \textit{integrated} growth
        rates for higher background temperature in Figure~\ref{fig2}c.}
    \label{fig7}
\end{figure*}

\cite{2007SoPh..241..127K} (TS07) found that growth rates decrease with
increasing gyro-harmonic number. They performed calculations for a power-law
distribution (power-law index $\delta = 10$) for the loss-cone angle
$30^\circ$. Their approach was to estimate the growth rates only at a specific
part of the resonance ellipse (Equation~\ref{eq1}). In contrast, in our approach
we integrate over the whole ellipse in Equation~\ref{eq-growth} for each
specific point in the $\omega-k_\perp$ space \citep{2005A&A...438..341K} (K05).
To obtain the integrated growth rate, we integrate over the UH dispersion
branch, which produces more reliable results according to the tests by
electromagnetic, relativistic particle-in-cell simulations
\citep{2019ApJ...881...21B,2021A&A...649A.145M} (BK19). Though TS07 used a
different kind of loss-cone distribution, the decrease of the growth rates with
increasing gyro-harmonic number for the loss-cone angle 30$^\circ$ is in
agreement with our results. Nonetheless, the exact growth rate values differ.

We compare these three approaches of growth rate calculations in
Figure~\ref{fig7} for the gyro-harmonic number $s \sim 50$ as a
function of the loss-cone angle. In all three cases, the growth rates are
selected on the UH dispersion branch for 0.5~MK and $v_\mathrm{\kappa} =
0.3\,c$. The maximal growth rates of the K05 method are smaller than those of the KS07
method for loss-cone angles $\leq$65$^\circ$. However, growth rates for the angle
80$^\circ$, which are smaller in the KT07 method, are the largest in the K05
method. Moreover, all growth rate profiles are relatively smoother for the  KT07
method than for the K05 method. The integrated growth rates of the BK19 method are
systematically lower than the maximal growth rates (from the definition of the
average along the dispersion branch in Equation~\ref{eq:integrated}). In
agreement with BK19, who studied the integrated growth rates only for small
gyro-harmonic numbers $\omega_\mathrm{pe}/ \omega_\mathrm{ce} \approx 4-5.2$,
the peaks of the \textit{integrated} growth rate are more distinct than
(normal) growth rates also for high gyro-harmonic numbers.

\citet{2019SoPh..294...29Y} among others computed growth rates for the loss-cone
angle 80$^\circ$ using the same method as KS07. They found negative growth
rates for large loss-cone angles. However, they applied a loss-cone
$\kappa$-distribution with an additional cut-off at $v_\mathrm{m}=0.3\,c$,
where $v_\mathrm{m}$ is the cut-off velocity.
For distributions without the cut-off,
they found that the growth rates are very broad for
$\theta_\mathrm{c} \sim 50^\circ$. As we show in Figure~\ref{fig7}, the
integrated growth rate peaks may still be distinct.

Increasing the $\kappa$ index of the
$\kappa$-dis\-tri\-bu\-tion from two to one hundred, the integrated growth rates increase,
and their maxima become more
distinct. Increasing $\kappa$
index, there are more particles with smaller $\kappa$-distribution velocities.
Therefore, these particles increase the distribution gradients and contribute
more in the integral of Equation~\ref{eq-growth} over the resonance ellipse.

We conclude that the DPR instability model may produce solar radio zebras with
high gyro-harmonic numbers for large loss-cone angles $\sim80^\circ$. The
growth rate maxima are distinct, and they are high enough to generate
zebra patterns.

\begin{acks}[Acknowledgements]
We acknowledge the valuable comments by an anonymous reviewer and the proof reading
by Dr. Patricio A. Mu\~{n}oz.
\end{acks}
\begin{fundinginformation}
	J. B. acknowledges support from the German Science Foundation (DFG) project BU
	777-17-1. M. K. acknowledges support from the project RVO-67985815 and GA\v{C}R
	grants 20-09922J, 20-07908S, 21-16508J, and 22-34841S.
	The authors gratefully acknowledge the Gauss Centre for Supercomputing e.V.
	(www.gauss-centre.eu) for partially funding this project by providing computing
	time on the GCS Su\-per\-com\-pu\-ter Su\-perMUC-NG at Leibniz Supercomputing Centre
	(www.lrz.de), project pn73ne. This work was also supported by the
	Ministry of Education, Youth and Sports of the Czech Republic through the e-INFRA CZ (ID:90140).
\end{fundinginformation}




\begin{dataavailability}
The datasets generated during and/or analyzed during the current study are available
from the corresponding author on reasonable request.
\end{dataavailability}

\begin{conflict}
The authors declare that there are no conflicts of interest.
\end{conflict}

\bibliographystyle{spr-mp-sola}
\bibliography{benacek}

\begin{thebibliography}{48}
\ifx\bisbn     \undefined \def\bisbn  #1{ISBN #1}\fi
\ifx\binits    \undefined \def\binits#1{#1}\fi
\ifx\bauthor   \undefined \def\bauthor#1{#1}\fi
\ifx\batitle   \undefined \def\batitle#1{#1}\fi
\ifx\bjtitle   \undefined \def\bjtitle#1{\textit{#1}}\fi
\ifx\bvolume   \undefined \def\bvolume#1{\textbf{#1}}\fi
\ifx\byear     \undefined \def\byear#1{#1}\fi
\ifx\bissue    \undefined \def\bissue#1{#1}\fi
\ifx\bfpage    \undefined \def\bfpage#1{#1}\fi
\ifx\blpage    \undefined \def\blpage #1{#1}\fi
\ifx\burl      \undefined \def\burl#1{#1}\fi
\ifx\href      \undefined \def\href#1#2{#2}\fi
\ifx\betal     \undefined \def\betal{et al.}\fi
\ifx\bctitle   \undefined \def\bctitle#1{#1}\fi
\ifx\beditor   \undefined \def\beditor#1{#1}\fi
\ifx\bbtitle   \undefined \def\bbtitle#1{\textit{#1}}\fi
\ifx\bedition  \undefined \def\bedition#1{#1}\fi
\ifx\bseriesno \undefined \def\bseriesno#1{\textbf{#1}}\fi
\ifx\blocation \undefined \def\blocation#1{#1}\fi
\ifx\bsertitle \undefined \def\bsertitle#1{\textit{#1}}\fi
\ifx\bsnm      \undefined \def\bsnm#1{#1}\fi
\ifx\bsuffix   \undefined \def\bsuffix#1{#1}\fi
\ifx\bparticle \undefined \def\bparticle#1{#1}\fi
\ifx\barticle  \undefined \def\barticle#1{}\fi
\ifx\binstitute  \undefined \def\binstitute#1{#1}\fi
\ifx\bpublisher  \undefined \def\bpublisher#1{#1}\fi
\ifx\doiurl    \undefined \def\doiurl#1{\href{#1}{DOI}}\fi
\makeatletter
\def\safeHref#1#2#3{\in@{http}{#2}\ifin@\href{#2}{#3}\else\href{#1#2}{#3}\fi}
\makeatother
\ifx\adsurl    \undefined
  \def\adsurl#1{\safeHref{https://ui.adsabs.harvard.edu/abs/}{#1}{ADS}}\fi
\ifx\arxivurl  \undefined
  \def\arxivurl#1{\safeHref{http://arxiv.org/abs/}{#1}{arXiv}}\fi
\ifx\botherref \undefined \def\botherref#1{}\fi
\ifx\url       \undefined \def\url#1{#1}\fi
\ifx\bchapter  \undefined \def\bchapter#1{}\fi
\ifx\bbook     \undefined \def\bbook#1{}\fi
\ifx\bcomment  \undefined \def\bcomment#1{#1}\fi
\ifx\oauthor   \undefined \def\oauthor#1{#1}\fi
\ifx\citeauthoryear \undefined\def \citeauthoryear#1{#1}\fi
\def\endbibitem {}
\ifx\bconflocation  \undefined \def\bconflocation#1{#1} \fi

\bibitem[\protect\citeauthoryear{{Aschwanden}}{2004}]{2004psci.book.....A}
\begin{bbook}
\bauthor{\bsnm{{Aschwanden}}, \binits{M.J.}}:
\byear{2004},
\bbtitle{{Physics of the Solar Corona. An Introduction}}.
\adsurl{2004psci.book.....A}.
\end{bbook}
\endbibitem

\bibitem[\protect\citeauthoryear{{B{\'a}rta} and
  {Karlick{\'y}}}{2006}]{2006A&A...450..359B}
\begin{barticle}
\bauthor{\bsnm{{B{\'a}rta}}, \binits{M.}},
\bauthor{\bsnm{{Karlick{\'y}}}, \binits{M.}}:
\byear{2006},
\batitle{{Interference patterns in solar radio spectra: high-resolution
  structural analysis of the corona}}.
\bjtitle{\aap}
\bvolume{450},
\bfpage{359}.
\doiurl{https://doi.org/10.1051/0004-6361:20054386}.
\adsurl{2006A&A...450..359B}.
\end{barticle}
\endbibitem

\bibitem[\protect\citeauthoryear{{Ben{\'a}{\v{c}}ek} and
  {Karlick{\'y}}}{2018}]{2018A&A...611A..60B}
\begin{barticle}
\bauthor{\bsnm{{Ben{\'a}{\v{c}}ek}}, \binits{J.}},
\bauthor{\bsnm{{Karlick{\'y}}}, \binits{M.}}:
\byear{2018},
\batitle{{Double plasma resonance instability as a source of solar zebra
  emission}}.
\bjtitle{\aap}
\bvolume{611},
\bfpage{A60}.
\doiurl{https://doi.org/10.1051/0004-6361/201731424}.
\adsurl{2018A&A...611A..60B}.
\end{barticle}
\endbibitem

\bibitem[\protect\citeauthoryear{{Ben{\'a}{\v{c}}ek} and
  {Karlick{\'y}}}{2019}]{2019ApJ...881...21B}
\begin{barticle}
\bauthor{\bsnm{{Ben{\'a}{\v{c}}ek}}, \binits{J.}},
\bauthor{\bsnm{{Karlick{\'y}}}, \binits{M.}}:
\byear{2019},
\batitle{{Growth Rates of the Electrostatic Waves in Radio Zebra Models}}.
\bjtitle{\apj}
\bvolume{881},
\bfpage{21}.
\doiurl{https://doi.org/10.3847/1538-4357/ab2bfc}.
\adsurl{2019ApJ...881...21B}.
\end{barticle}
\endbibitem

\bibitem[\protect\citeauthoryear{{Ben{\'a}{\v{c}}ek}, {Karlick{\'y}}, and
  {Yasnov}}{2017}]{2017A&A...598A.106B}
\begin{barticle}
\bauthor{\bsnm{{Ben{\'a}{\v{c}}ek}}, \binits{J.}},
\bauthor{\bsnm{{Karlick{\'y}}}, \binits{M.}},
\bauthor{\bsnm{{Yasnov}}, \binits{L.V.}}:
\byear{2017},
\batitle{{Temperature dependent growth rates of the upper-hybrid waves and
  solar radio zebra patterns}}.
\bjtitle{\aap}
\bvolume{598},
\bfpage{A106}.
\doiurl{https://doi.org/10.1051/0004-6361/201629717}.
\adsurl{2017A&A...598A.106B}.
\end{barticle}
\endbibitem

\bibitem[\protect\citeauthoryear{{Chen} et~al.}{2011}]{2011ApJ...736...64C}
\begin{barticle}
\bauthor{\bsnm{{Chen}}, \binits{B.}},
\bauthor{\bsnm{{Bastian}}, \binits{T.S.}},
\bauthor{\bsnm{{Gary}}, \binits{D.E.}},
\bauthor{\bsnm{{Jing}}, \binits{J.}}:
\byear{2011},
\batitle{{Spatially and Spectrally Resolved Observations of a Zebra Pattern in
  a Solar Decimetric Radio Burst}}.
\bjtitle{\apj}
\bvolume{736},
\bfpage{64}.
\doiurl{https://doi.org/10.1088/0004-637X/736/1/64}.
\adsurl{2011ApJ...736...64C}.
\end{barticle}
\endbibitem

\bibitem[\protect\citeauthoryear{{Chernov}}{2011}]{2011fssr.book.....C}
\begin{bbook}
\bauthor{\bsnm{{Chernov}}, \binits{G.}}:
\byear{2011},
\bbtitle{{Fine Structure of Solar Radio Bursts}}.
\adsurl{2011fssr.book.....C}.
\end{bbook}
\endbibitem

\bibitem[\protect\citeauthoryear{{Chernov}, {Formichev}, and
  {Fainshtein}}{}]{Chernov2020}
\begin{botherref}
\oauthor{\bsnm{{Chernov}}, \binits{G.}},
\oauthor{\bsnm{{Formichev}}, \binits{V.}},
\oauthor{\bsnm{{Fainshtein}}, \binits{S.}}:
\textit{Resent Results on the fine structure in Cosmic radio Emission},
Mauritius: Lambert Academic Publishing.
\end{botherref}
\endbibitem

\bibitem[\protect\citeauthoryear{{Chernov}}{1976}]{1976SvA....20..449C}
\begin{barticle}
\bauthor{\bsnm{{Chernov}}, \binits{G.P.}}:
\byear{1976},
\batitle{{Microstructure in the continuous radiation of type IV meter bursts.
  Observations and model of the source}}.
\bjtitle{\sovast}
\bvolume{20},
\bfpage{449}.
\adsurl{1976SvA....20..449C}.
\end{barticle}
\endbibitem

\bibitem[\protect\citeauthoryear{{Chernov}}{1990}]{1990SoPh..130...75C}
\begin{barticle}
\bauthor{\bsnm{{Chernov}}, \binits{G.P.}}:
\byear{1990},
\batitle{{Whistlers in the Solar Corona and Their Relevance to Fine Structures
  of Type-Iv Radio Emission}}.
\bjtitle{\solphys}
\bvolume{130},
\bfpage{75}.
\doiurl{https://doi.org/10.1007/BF00156780}.
\adsurl{1990SoPh..130...75C}.
\end{barticle}
\endbibitem

\bibitem[\protect\citeauthoryear{{Chernov}}{2006}]{2006SSRv..127..195C}
\begin{barticle}
\bauthor{\bsnm{{Chernov}}, \binits{G.P.}}:
\byear{2006},
\batitle{{Solar Radio Bursts with Drifting Stripes in Emission and
  Absorption}}.
\bjtitle{\ssr}
\bvolume{127},
\bfpage{195}.
\doiurl{https://doi.org/10.1007/s11214-006-9141-7}.
\adsurl{2006SSRv..127..195C}.
\end{barticle}
\endbibitem

\bibitem[\protect\citeauthoryear{{Chernov} et~al.}{2005}]{2005A&A...437.1047C}
\begin{barticle}
\bauthor{\bsnm{{Chernov}}, \binits{G.P.}},
\bauthor{\bsnm{{Yan}}, \binits{Y.H.}},
\bauthor{\bsnm{{Fu}}, \binits{Q.J.}},
\bauthor{\bsnm{{Tan}}, \binits{C.M.}}:
\byear{2005},
\batitle{{Recent data on zebra patterns}}.
\bjtitle{\aap}
\bvolume{437},
\bfpage{1047}.
\doiurl{https://doi.org/10.1051/0004-6361:20042578}.
\adsurl{2005A&A...437.1047C}.
\end{barticle}
\endbibitem

\bibitem[\protect\citeauthoryear{{Ji{\v{r}}i{\v{c}}ka} and
  {Karlick{\'y}}}{2008}]{2008SoPh..253...95J}
\begin{barticle}
\bauthor{\bsnm{{Ji{\v{r}}i{\v{c}}ka}}, \binits{K.}},
\bauthor{\bsnm{{Karlick{\'y}}}, \binits{M.}}:
\byear{2008},
\batitle{{Narrowband Pulsating Decimeter Structure Observed by the New
  Ond{\v{r}}ejov Solar Radio Spectrograph}}.
\bjtitle{\solphys}
\bvolume{253},
\bfpage{95}.
\doiurl{https://doi.org/10.1007/s11207-008-9118-7}.
\adsurl{2008SoPh..253...95J}.
\end{barticle}
\endbibitem

\bibitem[\protect\citeauthoryear{{Karlick{\'y}}}{2013}]{2013A&A...552A..90K}
\begin{barticle}
\bauthor{\bsnm{{Karlick{\'y}}}, \binits{M.}}:
\byear{2013},
\batitle{{Radio continua modulated by waves: Zebra patterns in solar and pulsar
  radio spectra?}}
\bjtitle{\aap}
\bvolume{552},
\bfpage{A90}.
\doiurl{https://doi.org/10.1051/0004-6361/201321356}.
\adsurl{2013A&A...552A..90K}.
\end{barticle}
\endbibitem

\bibitem[\protect\citeauthoryear{{Karlick{\'y}} and
  {Kosugi}}{2004}]{2004A&A...419.1159K}
\begin{barticle}
\bauthor{\bsnm{{Karlick{\'y}}}, \binits{M.}},
\bauthor{\bsnm{{Kosugi}}, \binits{T.}}:
\byear{2004},
\batitle{{Acceleration and heating processes in a collapsing magnetic trap}}.
\bjtitle{\aap}
\bvolume{419},
\bfpage{1159}.
\doiurl{https://doi.org/10.1051/0004-6361:20034323}.
\adsurl{2004A&A...419.1159K}.
\end{barticle}
\endbibitem

\bibitem[\protect\citeauthoryear{{Karlick{\'y}} and
  {Yasnov}}{2015}]{2015A&A...581A.115K}
\begin{barticle}
\bauthor{\bsnm{{Karlick{\'y}}}, \binits{M.}},
\bauthor{\bsnm{{Yasnov}}, \binits{L.V.}}:
\byear{2015},
\batitle{{Determination of plasma parameters in solar zebra radio sources}}.
\bjtitle{\aap}
\bvolume{581},
\bfpage{A115}.
\doiurl{https://doi.org/10.1051/0004-6361/201526785}.
\adsurl{2015A&A...581A.115K}.
\end{barticle}
\endbibitem

\bibitem[\protect\citeauthoryear{{Karlick{\'y}} and
  {Yasnov}}{2018}]{2018ApJ...867...28K}
\begin{barticle}
\bauthor{\bsnm{{Karlick{\'y}}}, \binits{M.}},
\bauthor{\bsnm{{Yasnov}}, \binits{L.V.}}:
\byear{2018},
\batitle{{Determination of Plasma Parameters in Radio Sources of Solar
  Zebra-patterns Based on Relations between the Zebra-stripe Frequencies and
  Gyro-harmonic Numbers}}.
\bjtitle{\apj}
\bvolume{867},
\bfpage{28}.
\doiurl{https://doi.org/10.3847/1538-4357/aae1f8}.
\adsurl{2018ApJ...867...28K}.
\end{barticle}
\endbibitem

\bibitem[\protect\citeauthoryear{{Karlick{\'y}} and
  {Yasnov}}{2021}]{2021A&A...646A.179K}
\begin{barticle}
\bauthor{\bsnm{{Karlick{\'y}}}, \binits{M.}},
\bauthor{\bsnm{{Yasnov}}, \binits{L.V.}}:
\byear{2021},
\batitle{{Spatial quasi-periodic variations of the plasma density and magnetic
  field in zebra radio sources}}.
\bjtitle{\aap}
\bvolume{646},
\bfpage{A179}.
\doiurl{https://doi.org/10.1051/0004-6361/202039850}.
\adsurl{2021A&A...646A.179K}.
\end{barticle}
\endbibitem

\bibitem[\protect\citeauthoryear{{Klimchuk}}{2000}]{2000SoPh..193...53K}
\begin{barticle}
\bauthor{\bsnm{{Klimchuk}}, \binits{J.A.}}:
\byear{2000},
\batitle{{Cross-Sectional Properties of Coronal Loops}}.
\bjtitle{\solphys}
\bvolume{193},
\bfpage{53}.
\doiurl{https://doi.org/10.1023/A:1005210127703}.
\adsurl{2000SoPh..193...53K}.
\end{barticle}
\endbibitem

\bibitem[\protect\citeauthoryear{{Krucker}, {Masuda}, and
  {White}}{2020}]{2020ApJ...894..158K}
\begin{barticle}
\bauthor{\bsnm{{Krucker}}, \binits{S.}},
\bauthor{\bsnm{{Masuda}}, \binits{S.}},
\bauthor{\bsnm{{White}}, \binits{S.M.}}:
\byear{2020},
\batitle{{Microwave and Hard X-Ray Flare Observations by NoRH/NoRP and RHESSI:
  Peak-flux Correlations}}.
\bjtitle{\apj}
\bvolume{894},
\bfpage{158}.
\doiurl{https://doi.org/10.3847/1538-4357/ab8644}.
\adsurl{2020ApJ...894..158K}.
\end{barticle}
\endbibitem

\bibitem[\protect\citeauthoryear{{Kuijpers}}{1980}]{1980IAUS...86..341K}
\begin{bchapter}
\bauthor{\bsnm{{Kuijpers}}, \binits{J.}}:
\byear{1980},
\bctitle{{Theory of type IV DM bursts}}.
In: \beditor{\bsnm{{Kundu}}, \binits{M.R.}},
\beditor{\bsnm{{Gergely}}, \binits{T.E.}} (eds.)
\bbtitle{Radio Physics of the Sun}
\bseriesno{86},
\bfpage{341}.
\doiurl{https://doi.org/10.1017/S0074180900037098}.
\adsurl{1980IAUS...86..341K}.
\end{bchapter}
\endbibitem

\bibitem[\protect\citeauthoryear{{Kuijpers}}{1975}]{1975PhDT.........1K}
\begin{botherref}
\oauthor{\bsnm{{Kuijpers}}, \binits{J.M.E.}}:
1975,
{Collective wave-particle interactions in solar type IV radio sources}.
PhD thesis,
-.
\adsurl{1975PhDT.........1K}.
\end{botherref}
\endbibitem

\bibitem[\protect\citeauthoryear{{Kuznetsov}}{2005}]{2005A&A...438..341K}
\begin{barticle}
\bauthor{\bsnm{{Kuznetsov}}, \binits{A.A.}}:
\byear{2005},
\batitle{{Generation of microwave bursts with zebra pattern by nonlinear
  interaction of Bernstein modes}}.
\bjtitle{\aap}
\bvolume{438},
\bfpage{341}.
\doiurl{https://doi.org/10.1051/0004-6361:20052712}.
\adsurl{2005A&A...438..341K}.
\end{barticle}
\endbibitem

\bibitem[\protect\citeauthoryear{{Kuznetsov} and
  {Tsap}}{2007}]{2007SoPh..241..127K}
\begin{barticle}
\bauthor{\bsnm{{Kuznetsov}}, \binits{A.A.}},
\bauthor{\bsnm{{Tsap}}, \binits{Y.T.}}:
\byear{2007},
\batitle{{Loss-Cone Instability and Formation of Zebra Patterns in Type IV
  Solar Radio Bursts}}.
\bjtitle{\solphys}
\bvolume{241},
\bfpage{127}.
\doiurl{https://doi.org/10.1007/s11207-006-0351-7}.
\adsurl{2007SoPh..241..127K}.
\end{barticle}
\endbibitem

\bibitem[\protect\citeauthoryear{{LaBelle} et~al.}{2003}]{2003ApJ...593.1195L}
\begin{barticle}
\bauthor{\bsnm{{LaBelle}}, \binits{J.}},
\bauthor{\bsnm{{Treumann}}, \binits{R.A.}},
\bauthor{\bsnm{{Yoon}}, \binits{P.H.}},
\bauthor{\bsnm{{Karlicky}}, \binits{M.}}:
\byear{2003},
\batitle{{A Model of Zebra Emission in Solar Type IV Radio Bursts}}.
\bjtitle{\apj}
\bvolume{593},
\bfpage{1195}.
\doiurl{https://doi.org/10.1086/376732}.
\adsurl{2003ApJ...593.1195L}.
\end{barticle}
\endbibitem

\bibitem[\protect\citeauthoryear{{Laptukhov} and
  {Chernov}}{2009}]{2009PlPhR..35..160L}
\begin{barticle}
\bauthor{\bsnm{{Laptukhov}}, \binits{A.I.}},
\bauthor{\bsnm{{Chernov}}, \binits{G.P.}}:
\byear{2009},
\batitle{{Concerning mechanisms for the zebra pattern formation in the solar
  radio emission}}.
\bjtitle{Plasma Physics Reports}
\bvolume{35},
\bfpage{160}.
\doiurl{https://doi.org/10.1134/S1063780X09020081}.
\adsurl{2009PlPhR..35..160L}.
\end{barticle}
\endbibitem

\bibitem[\protect\citeauthoryear{{Ledenev}, {Yan}, and
  {Fu}}{2006}]{2006SoPh..233..129L}
\begin{barticle}
\bauthor{\bsnm{{Ledenev}}, \binits{V.G.}},
\bauthor{\bsnm{{Yan}}, \binits{Y.}},
\bauthor{\bsnm{{Fu}}, \binits{Q.}}:
\byear{2006},
\batitle{{Interference Mechanism of ``Zebra-Pattern'' Formation in Solar Radio
  Emission}}.
\bjtitle{\solphys}
\bvolume{233},
\bfpage{129}.
\doiurl{https://doi.org/10.1007/s11207-006-2099-5}.
\adsurl{2006SoPh..233..129L}.
\end{barticle}
\endbibitem

\bibitem[\protect\citeauthoryear{{Lee} et~al.}{2018}]{2018JGRA..123.7320L}
\begin{barticle}
\bauthor{\bsnm{{Lee}}, \binits{J.}},
\bauthor{\bsnm{{Yoon}}, \binits{P.H.}},
\bauthor{\bsnm{{Seough}}, \binits{J.}},
\bauthor{\bsnm{{L{\'o}pez}}, \binits{R.A.}},
\bauthor{\bsnm{{Hwang}}, \binits{J.}},
\bauthor{\bsnm{{Lee}}, \binits{J.}},
\bauthor{\bsnm{{Choe}}, \binits{G.S.}}:
\byear{2018},
\batitle{{Simulation and Quasi-linear Theory of Magnetospheric Bernstein Mode
  Instability}}.
\bjtitle{Journal of Geophysical Research (Space Physics)}
\bvolume{123},
\bfpage{7320}.
\doiurl{https://doi.org/10.1029/2018JA025667}.
\adsurl{2018JGRA..123.7320L}.
\end{barticle}
\endbibitem

\bibitem[\protect\citeauthoryear{Li
  et~al.}{2021}]{Li_Chen_Ni_Tan_Ning_Zhang_2021}
\begin{barticle}
\bauthor{\bsnm{Li}, \binits{C.}},
\bauthor{\bsnm{Chen}, \binits{Y.}},
\bauthor{\bsnm{Ni}, \binits{S.}},
\bauthor{\bsnm{Tan}, \binits{B.}},
\bauthor{\bsnm{Ning}, \binits{H.}},
\bauthor{\bsnm{Zhang}, \binits{Z.}}:
\byear{2021},
\batitle{{PIC} Simulation of Double Plasma Resonance and Zebra Pattern of Solar
  Radio Bursts}.
\bjtitle{The Astrophysical Journal Letters}
\bvolume{909},
\bfpage{L5}.
\doiurl{https://doi.org/10.3847/2041-8213/abe708}.
\burl{https://doi.org/10.3847/2041-8213/abe708}.
\end{barticle}
\endbibitem

\bibitem[\protect\citeauthoryear{{Livadiotis} and
  {McComas}}{2013}]{2013SSRv..175..183L}
\begin{barticle}
\bauthor{\bsnm{{Livadiotis}}, \binits{G.}},
\bauthor{\bsnm{{McComas}}, \binits{D.J.}}:
\byear{2013},
\batitle{{Understanding Kappa Distributions: A Toolbox for Space Science and
  Astrophysics}}.
\bjtitle{\ssr}
\bvolume{175},
\bfpage{183}.
\doiurl{https://doi.org/10.1007/s11214-013-9982-9}.
\adsurl{2013SSRv..175..183L}.
\end{barticle}
\endbibitem

\bibitem[\protect\citeauthoryear{{L{\"o}rin{\v{c}}{\'\i}k}
  et~al.}{2020}]{2020ApJ...893...34L}
\begin{barticle}
\bauthor{\bsnm{{L{\"o}rin{\v{c}}{\'\i}k}}, \binits{J.}},
\bauthor{\bsnm{{Dud{\'\i}k}}, \binits{J.}},
\bauthor{\bsnm{{del Zanna}}, \binits{G.}},
\bauthor{\bsnm{{Dzif{\v{c}}{\'a}kov{\'a}}}, \binits{E.}},
\bauthor{\bsnm{{Mason}}, \binits{H.E.}}:
\byear{2020},
\batitle{{Plasma Diagnostics from Active Region and Quiet-Sun Spectra Observed
  by Hinode/EIS: Quantifying the Departures from a Maxwellian Distribution}}.
\bjtitle{\apj}
\bvolume{893},
\bfpage{34}.
\doiurl{https://doi.org/10.3847/1538-4357/ab8010}.
\adsurl{2020ApJ...893...34L}.
\end{barticle}
\endbibitem

\bibitem[\protect\citeauthoryear{{Manthei} et~al.}{2021}]{2021A&A...649A.145M}
\begin{barticle}
\bauthor{\bsnm{{Manthei}}, \binits{A.C.}},
\bauthor{\bsnm{{Ben{\'a}{\v{c}}ek}}, \binits{J.}},
\bauthor{\bsnm{{Mu{\~n}oz}}, \binits{P.A.}},
\bauthor{\bsnm{{B{\"u}chner}}, \binits{J.}}:
\byear{2021},
\batitle{{Refining pulsar radio emission due to streaming instabilities: Linear
  theory and PIC simulations in a wide parameter range}}.
\bjtitle{\aap}
\bvolume{649},
\bfpage{A145}.
\doiurl{https://doi.org/10.1051/0004-6361/202039907}.
\adsurl{2021A&A...649A.145M}.
\end{barticle}
\endbibitem

\bibitem[\protect\citeauthoryear{{Melrose}}{1986}]{Melrose1986}
\begin{bbook}
\bauthor{\bsnm{{Melrose}}, \binits{D.B.}}:
\byear{1986},
\bbtitle{{Instabilities in Space and Laboratory Plasmas}}.
\adsurl{1986islp.book.....M}.
\end{bbook}
\endbibitem

\bibitem[\protect\citeauthoryear{{Mollwo}}{1983}]{1983SoPh...83..305M}
\begin{barticle}
\bauthor{\bsnm{{Mollwo}}, \binits{L.}}:
\byear{1983},
\batitle{{Interpretation of Patterns of Drifting ZEBRA Stripes}}.
\bjtitle{\solphys}
\bvolume{83},
\bfpage{305}.
\doiurl{https://doi.org/10.1007/BF00148282}.
\adsurl{1983SoPh...83..305M}.
\end{barticle}
\endbibitem

\bibitem[\protect\citeauthoryear{{Mollwo}}{1988}]{1988SoPh..116..323M}
\begin{barticle}
\bauthor{\bsnm{{Mollwo}}, \binits{L.}}:
\byear{1988},
\batitle{{The Magneto-Hydrostatic Field in the Region of ZEBRA Patterns in
  Solar Type-Iv Dm-Bursts}}.
\bjtitle{\solphys}
\bvolume{116},
\bfpage{323}.
\doiurl{https://doi.org/10.1007/BF00157482}.
\adsurl{1988SoPh..116..323M}.
\end{barticle}
\endbibitem

\bibitem[\protect\citeauthoryear{{Ni} et~al.}{2020}]{Ni_2020}
\begin{barticle}
\bauthor{\bsnm{{Ni}}, \binits{S.}},
\bauthor{\bsnm{{Chen}}, \binits{Y.}},
\bauthor{\bsnm{{Li}}, \binits{C.}},
\bauthor{\bsnm{{Zhang}}, \binits{Z.}},
\bauthor{\bsnm{{Ning}}, \binits{H.}},
\bauthor{\bsnm{{Kong}}, \binits{X.}},
\bauthor{\bsnm{{Wang}}, \binits{B.}},
\bauthor{\bsnm{{Hosseinpour}}, \binits{M.}}:
\byear{2020},
\batitle{{Plasma Emission Induced by Electron Cyclotron Maser Instability in
  Solar Plasmas with a Large Ratio of Plasma Frequency to Gyrofrequency}}.
\bjtitle{\apjl}
\bvolume{891},
\bfpage{L25}.
\doiurl{https://doi.org/10.3847/2041-8213/ab7750}.
\adsurl{2020ApJ...891L..25N}.
\end{barticle}
\endbibitem

\bibitem[\protect\citeauthoryear{{Rosenberg}}{1972}]{1972SoPh...25..188R}
\begin{barticle}
\bauthor{\bsnm{{Rosenberg}}, \binits{H.}}:
\byear{1972},
\batitle{{A Possibly Direct Measurement of Coronal Magnetic Field Strengths}}.
\bjtitle{\solphys}
\bvolume{25},
\bfpage{188}.
\doiurl{https://doi.org/10.1007/BF00155756}.
\adsurl{1972SoPh...25..188R}.
\end{barticle}
\endbibitem

\bibitem[\protect\citeauthoryear{Slottje}{Apr 1982}]{slo81}
\begin{botherref}
\oauthor{\bsnm{Slottje}, \binits{C.}}:
Apr 1982,
\textit{Atlas of fine structures of dynamic spectra of solar type IV-dm and
  some type II radio bursts}.
\url{http://inis.iaea.org/search/search.aspx?orig_q=RN:13712618}.
\end{botherref}
\endbibitem

\bibitem[\protect\citeauthoryear{{Tan}}{2010}]{2010Ap&SS.325..251T}
\begin{barticle}
\bauthor{\bsnm{{Tan}}, \binits{B.}}:
\byear{2010},
\batitle{{A physical explanation of solar microwave Zebra pattern with the
  current-carrying plasma loop model}}.
\bjtitle{\apss}
\bvolume{325},
\bfpage{251}.
\doiurl{https://doi.org/10.1007/s10509-009-0193-5}.
\adsurl{2010Ap&SS.325..251T}.
\end{barticle}
\endbibitem

\bibitem[\protect\citeauthoryear{{Tan} et~al.}{2012}]{2012ApJ...744..166T}
\begin{barticle}
\bauthor{\bsnm{{Tan}}, \binits{B.}},
\bauthor{\bsnm{{Yan}}, \binits{Y.}},
\bauthor{\bsnm{{Tan}}, \binits{C.}},
\bauthor{\bsnm{{Sych}}, \binits{R.}},
\bauthor{\bsnm{{Gao}}, \binits{G.}}:
\byear{2012},
\batitle{{Microwave Zebra Pattern Structures in the X2.2 Solar Flare on 2011
  February 15}}.
\bjtitle{\apj}
\bvolume{744},
\bfpage{166}.
\doiurl{https://doi.org/10.1088/0004-637X/744/2/166}.
\adsurl{2012ApJ...744..166T}.
\end{barticle}
\endbibitem

\bibitem[\protect\citeauthoryear{{Tan} et~al.}{2014}]{2014ApJ...780..129T}
\begin{barticle}
\bauthor{\bsnm{{Tan}}, \binits{B.}},
\bauthor{\bsnm{{Tan}}, \binits{C.}},
\bauthor{\bsnm{{Zhang}}, \binits{Y.}},
\bauthor{\bsnm{{M{\'e}sz{\'a}rosov{\'a}}}, \binits{H.}},
\bauthor{\bsnm{{Karlick{\'y}}}, \binits{M.}}:
\byear{2014},
\batitle{{Statistics and Classification of the Microwave Zebra Patterns
  Associated with Solar Flares}}.
\bjtitle{\apj}
\bvolume{780},
\bfpage{129}.
\doiurl{https://doi.org/10.1088/0004-637X/780/2/129}.
\adsurl{2014ApJ...780..129T}.
\end{barticle}
\endbibitem

\bibitem[\protect\citeauthoryear{{{\v{S}}vestka}
  et~al.}{1987}]{1987SoPh..108..237S}
\begin{barticle}
\bauthor{\bsnm{{{\v{S}}vestka}}, \binits{Z.F.}},
\bauthor{\bsnm{{Fontenla}}, \binits{J.M.}},
\bauthor{\bsnm{{Machado}}, \binits{M.E.}},
\bauthor{\bsnm{{Martin}}, \binits{S.F.}},
\bauthor{\bsnm{{Neidig}}, \binits{D.F.}},
\bauthor{\bsnm{{Poletto}}, \binits{G.}}:
\byear{1987},
\batitle{{Multi-thermal observations of newly formed loops in a dynamic
  flare}}.
\bjtitle{\solphys}
\bvolume{108},
\bfpage{237}.
\doiurl{https://doi.org/10.1007/BF00214164}.
\adsurl{1987SoPh..108..237S}.
\end{barticle}
\endbibitem

\bibitem[\protect\citeauthoryear{{Winglee} and
  {Dulk}}{1986}]{1986ApJ...307..808W}
\begin{barticle}
\bauthor{\bsnm{{Winglee}}, \binits{R.M.}},
\bauthor{\bsnm{{Dulk}}, \binits{G.A.}}:
\byear{1986},
\batitle{{The Electron-Cyclotron Maser Instability as a Source of Plasma
  Radiation}}.
\bjtitle{\apj}
\bvolume{307},
\bfpage{808}.
\doiurl{https://doi.org/10.1086/164467}.
\adsurl{1986ApJ...307..808W}.
\end{barticle}
\endbibitem

\bibitem[\protect\citeauthoryear{{Yasnov}}{2021}]{2021SoPh..296..139Y}
\begin{barticle}
\bauthor{\bsnm{{Yasnov}}, \binits{L.V.}}:
\byear{2021},
\batitle{{On the Magnetoacoustic Waves and Physical Conditions in Zebra Radio
  Sources}}.
\bjtitle{\solphys}
\bvolume{296},
\bfpage{139}.
\doiurl{https://doi.org/10.1007/s11207-021-01886-2}.
\adsurl{2021SoPh..296..139Y}.
\end{barticle}
\endbibitem

\bibitem[\protect\citeauthoryear{{Yasnov} and
  {Karlick{\'y}}}{2020}]{2020SoPh..295...96Y}
\begin{barticle}
\bauthor{\bsnm{{Yasnov}}, \binits{L.V.}},
\bauthor{\bsnm{{Karlick{\'y}}}, \binits{M.}}:
\byear{2020},
\batitle{{Magnetic Field, Electron Density and Their Spatial Scales in Zebra
  Pattern Radio Sources}}.
\bjtitle{\solphys}
\bvolume{295},
\bfpage{96}.
\doiurl{https://doi.org/10.1007/s11207-020-01652-w}.
\adsurl{2020SoPh..295...96Y}.
\end{barticle}
\endbibitem

\bibitem[\protect\citeauthoryear{{Yasnov}, {Ben{\'a}{\v{c}}ek}, and
  {Karlick{\'y}}}{2019}]{2019SoPh..294...29Y}
\begin{barticle}
\bauthor{\bsnm{{Yasnov}}, \binits{L.V.}},
\bauthor{\bsnm{{Ben{\'a}{\v{c}}ek}}, \binits{J.}},
\bauthor{\bsnm{{Karlick{\'y}}}, \binits{M.}}:
\byear{2019},
\batitle{{Growth Rates of the Upper-Hybrid Waves for Power-Law and Kappa
  Distributions with a Loss-Cone Anisotropy}}.
\bjtitle{\solphys}
\bvolume{294},
\bfpage{29}.
\doiurl{https://doi.org/10.1007/s11207-019-1415-9}.
\adsurl{2019SoPh..294...29Y}.
\end{barticle}
\endbibitem

\bibitem[\protect\citeauthoryear{{Zheleznyakov} and
  {Zlotnik}}{1975}]{1975SoPh...44..461Z}
\begin{barticle}
\bauthor{\bsnm{{Zheleznyakov}}, \binits{V.V.}},
\bauthor{\bsnm{{Zlotnik}}, \binits{E.Y.}}:
\byear{1975},
\batitle{{Cyclotron wave instability in the corona and origin of solar radio
  emission with fine structure. III. Origin of zebra-pattern.}}
\bjtitle{\solphys}
\bvolume{44},
\bfpage{461}.
\doiurl{https://doi.org/10.1007/BF00153225}.
\adsurl{1975SoPh...44..461Z}.
\end{barticle}
\endbibitem

\bibitem[\protect\citeauthoryear{{Zlotnik}}{2013}]{2013SoPh..284..579Z}
\begin{barticle}
\bauthor{\bsnm{{Zlotnik}}, \binits{E.Y.}}:
\byear{2013},
\batitle{{Instability of Electrons Trapped by the Coronal Magnetic Field and
  Its Evidence in the Fine Structure (Zebra Pattern) of Solar Radio Spectra}}.
\bjtitle{\solphys}
\bvolume{284},
\bfpage{579}.
\doiurl{https://doi.org/10.1007/s11207-012-0151-1}.
\adsurl{2013SoPh..284..579Z}.
\end{barticle}
\endbibitem

\end{thebibliography}

\end{article}

\end{document}